\documentclass[aps,prd,showpacs,notitlepage,nofootinbib,superscriptaddress,floatfix,showkeys,twocolumn]{revtex4-1}
\usepackage{comment}
\usepackage{graphicx}
\usepackage{amsmath}
\usepackage{float}
\usepackage{subfigure}
\usepackage{tensor}
\usepackage{natbib}
\usepackage[usenames]{color}
\usepackage{amssymb}
\usepackage{multirow}
\usepackage{longtable}
\usepackage{lipsum, babel}
\usepackage{epstopdf}
\newcommand{\bea}{\begin{aligned}}
\newcommand{\eea}{\end{aligned}}
\newcommand{\be}{\begin{equation}}
\newcommand{\ee}{\end{equation}}

\newcommand{\nno}{\nonumber}
\newcommand{\bse}{\begin{subequations}}
\newcommand{\ese}{\end{subequations}}

\usepackage{hyperref}
\hypersetup{
     colorlinks   = true,
     citecolor    = red,
     linkcolor    = blue,
     urlcolor     = blue,
}

\newcommand{\bmm}{\begin{multline}}
\newcommand{\emm}{\end{multline}}

\numberwithin{equation}{section}
\begin{document}
\title{Inflaton phenomenology via reheating in light of primordial gravitational
waves and the latest BICEP/$\mathbf{Keck}$ data}
\author{Ayan Chakraborty}%
\email{chakrabo@iitg.ac.in}
\affiliation{Department of Physics, 
Indian Institute of Technology, Guwahati, Assam 781039, India}%
\author{Md Riajul Haque}
\email{riaj.0009@gmail.com}
\affiliation{Centre for Strings, Gravitation, and Cosmology, Department of Physics, Indian Institute of Technology
Madras, Chennai 600036, India}
\author{Debaprasad Maity}
\email{debu@iitg.ac.in}
\affiliation{Department of Physics, 
Indian Institute of Technology, Guwahati, Assam 781039, India}
\author{Rajesh Mondal}
\email{mrajesh@iitg.ac.in}
\affiliation{Department of Physics, 
Indian Institute of Technology, Guwahati, Assam 781039, India}
\pagenumbering{arabic}
\renewcommand{\thesection}{\arabic{section}}

\begin{abstract}
We are in the era of precision cosmology which offers us a unique opportunity to investigate beyond standard model physics. Toward this endeavor, inflaton is assumed to be a perfect new physics candidate. In this submission, we explore the phenomenological impact of the latest observation of PLANCK and BICEP/$Keck$ data on the physics of inflation. We particularly study three different models of inflation, namely $\alpha$-attractor E, T, and the minimal plateau model. We further consider two different post-inflationary reheating dynamics driven by inflaton decaying into bosons and fermions. Given the latest data in the inflationary $(n_s-r)$ plane, we derive detailed phenomenological constraints on different inflaton parameters and the associated physical quantities, such as inflationary $e$-folding number, $N_{ k}$, reheating temperatures $T_{\rm re}$. Apart from considering direct observational data, we further incorporate the bounds from  primordial gravitational waves (PGWs) and different theoretical constraints. Rather than in the laboratory, our results illustrate the potential of present and future cosmological observations to look for new physics in the sky. 
\end{abstract}

\maketitle
\newpage
\section{\textbf{Introduction}}
Over the last decade, increasingly precise measurements of cosmic microwave background (CMB)  have led to a new era of precision cosmology. Inflation is assumed to be a unique mechanism \cite{Guth:1982ec,Starobinsky:1982ee,Albrecht:1982mp} in the early universe, which, apart from solving the problems of the standard big bang, has very precise predictions of large-scale inhomogeneous fluctuations. During the course of subsequent evolution, those fluctuations are translated into CMB anisotropy \cite{Sofue:2000jx,CMB-S4:2016ple,Planck:2018jri}. Therefore, accurate measurement of CMB anisotropy would be of fundamental importance for establishing the early inflationary phase. Two fundamental observables of interest are the scalar spectral index $n_s$ and tensor to scalar ratio $r$, which are directly connected to the inflaton, responsible for the inflation. In the realm of observations, considering  the BICEP/$Keck$(BK) 15 program \cite{BICEP2:2018kqh}, the measurement yielded a constraint on the $r_{0.05}<{0.07}$ at $95 \%$ CL, where subscript $k_0=0.05 {\rm Mpc}^{-1}$ is associated with the pivot scale.   
However, the latest BICEP/$Keck$ 18 \cite{BICEP:2021xfz} result along with Planck 2018 \cite{Planck:2018jri} yields the strongest constraints in the $(n_s,r)$ plane with $r_{0.05}<{0.036}$ and scalar spectral index $0.958<n_s<0.975$ at $95\,\%$ confidence level for $r=0.004$. In this submission, we intend to explore the impact of these latest observations on the three classes of the plateau-type inflationary model.

In the minimal framework, the standard model Higgs is assumed to be the best candidate for the inflaton field. However, Higgs inflation \cite{Bezrukov:2007ep} requires nonminimal gravitational coupling, which later on reformulated as Higgs-Starobinsky inflation \cite{Ellis:2013nxa,Kallosh:2014laa}, generically predicts very low scalar to tensor ratio $r\sim 0.003$ along with $n_s \sim (0.955,\,0.965)$ within the observational limit. Further generalization of such a model into a bigger class was invented as $\alpha$-attractor E and T model, which can be obtained from a spontaneously broken conformal invariant theory \cite{Kallosh:2013hoa,Kallosh:2013yoa}. Another class of model which are also consistent with observation is dubbed as minimal inflaton model \cite{Maity:2019ltu}. The common feature of all these models is the plateau region in the large field limit, and that leads to quasi-de Sitter expansion consistent with observation. 

Inflation is not the end of the story, though. Reheating, the phase when the inflaton field transfers its energy into the standard model fields yielding the radiation-filled universe is also of great importance when it comes to understanding the inflaton's real nature. Reheating is generically characterized by reheating temperature $T_{\rm re}$ and equation of state $w_\phi$, which are directly related to the inflaton$(\phi)$-radiation coupling and its potential $V(\phi)$. To realize reheating, we further investigate in detail the reheating dynamics by solving  the appropriate set of Boltzmann equations considering two different decay channels of inflaton, $\phi \rightarrow{ \bar f}f$ (fermionic) and $\phi\rightarrow {b b}$ (bosonic) \cite{Garcia:2020wiy,Garcia:2020eof,Haque:2023yra}. Where, these massless scalars or fermionic decay products will be considered as radiation, and their temperature at the end of reheating represents by $T_{\rm re}$. Lack of direct observation leads to a wide possible range of these parameters, reheating temperature within $(10^{15}~ {\rm GeV}, T_{\rm BBN}=4~{\rm MeV}$), and equation of state within $(-1/3, 1)$. Where $T_{\rm BBN}$ stands for the temperature when big bang nucleosynthesis (BBN) occurs and the light elements form.  While observable $(n_s,r)$ typically encodes only the intrinsic nature of inflaton to the leading order, reheating encodes much more. Inflaton is naturally thought to be a part of beyond the standard model of particle physics. Therefore, from the model-building perspective inflaton field can play an outstanding role in constructing a unified framework of cosmology and particle physics \cite{Garcia:2020eof,Giudice:2000ex,Maity:2018dgy,Haque:2019prw,Haque:2020zco,Garcia:2020wiy,Mambrini:2021zpp,Haque:2021mab,Clery:2021bwz,Haque:2022kez,Haque:2023yra}. Towards this goal recently few studies have been done considering matter like reheating \cite{Ellis:2021kad,Drewes:2022nhu,Drewes:2023bbs}. Our earlier attempts were mostly focused on the inflaton-dark matter sector. In this paper, we explore phenomenological constraints on the parameters of inflaton potential and its couplings and derive the bound on reheating parameters in the light of aforesaid combined data of Planck18, BK18, and BAO, along with primordial gravitational waves. 

Primordial gravitational waves (PGWs) are one of the profound predictions of inflation \cite{Grishchuk:1974ny,Starobinsky:1979ty}. Because of weak coupling, it carries not only the imprints of its own origin through inflationary observable $r$ but also of the post-inflationary evolution, particularly the reheating phase. The evolution of PGWs during the reheating phase and CMB observation is observed to set severe constraints on the inflation and reheating parameters. Around the CMB pivot scale the PGW spectrum, $\Omega^{\rm k}_{\rm GW}$ is constrained by $r< 0.036$ with typical dimensionless amplitude $\Omega^{\rm k}_{\rm GW}\sim 10^{-18}$. However, PGW of sub-horizon scale during reheating develops a $w_{\phi}$-dependent spectral tilt $\Omega^{\rm k}_{\rm GW} \propto 10^{-18} k^{-n_{w_{\phi}}}$, with $n_{w_{\phi}}=2(1-3 w_{\phi})/(1+3w_{\phi})$, which greatly enhances the magnitude of $\Omega^{\rm k}_{\rm GW}$ for equation $w_{\phi}>1/3$. Such enhancement will be observed to set a lower limit on the reheating temperature, which we defined as $T_{\rm re}^{\rm GW}$ though BBN constraint on total GW amplitude $\Omega_{\rm GW} \,h^2\leq1.7\times 10^{-6}$ obtained from dataset Planck-2018 + BICEP2/$Keck$ array \cite{Clarke:2020bil}. To put final bounds on the inflaton parameters, we further take into account perturbative and  limits on the inflaton-radiation coupling.  

The paper is organized as follows: In Sec.\ref{sc2}, we have started our discussion by demonstrating the inflation model. In Sec.\ref{sc3}, we discuss in detail the reheating dynamics for two different decay channels. PGWs is known to be an interesting cosmological observable which can significantly restrict the possible lower limit of reheating temperature through BBN constraint. We discuss this in detail in Sec.\ref{sc4}, and show the impact of such a limit on the inflaton-radiation coupling. In Secs.\ref{sc5} and \ref{sc6}, we illustrate different theoretical constraints, particularly on the inflaton-radiation coupling and its potential impact on the inflaton phenomenological parameter space. In Sec.\ref{sc7}, we discuss in detail the resulting constraints on parameters of inflaton and associated inflationary, reheating observable. Finally, we conclude with some future directions.
\section{\textbf{Model of inflation}}\label{sc2}
As pointed out in the Introduction, we explore three different single-field inflation models and left multifield models for our future work.
The associated potentials are 
\be
\label{attractorpotential}
V(\phi)  \;=\; 
\begin{cases}
\Lambda^4\,\left[1-e^{-\sqrt{\frac{2}{3\,\alpha}}\phi/M_p}\right]^{n}\,, & {\rm E - model}\,,\\[10pt]
\Lambda^4\,\tanh^{n}\left(\frac{\phi}{\sqrt{6\alpha}\,M_p}\right)\,,&  {\rm T-model}\,, \\[10pt]
\Lambda^4\,\frac{\phi^n}{\phi^n+\phi_*^n},& {\rm Minimal -model}\,, \\[10pt]
\end{cases}
\ee
One particularly notices that for $n=2$, the minimal model boils down to radion gauge inflation \cite{rgi}. In the large field limit, all the potential becomes constant  $V(\phi) \simeq \Lambda^4$ setting the scale of inflation, and the typical value it assumes $\sim (10^{15},10^{16})$ GeV. 
The remaining parameters $(n,\alpha,\phi_*)$ parametrizing  the shape of the potential near their minimum. Here one important point is to note that $\alpha$ is a dimensionless parameter whereas $\phi_*$ is measured in $M_{\rm p}$ unit. Near the minimum at $\phi \simeq 0$, all the potential assumes the form $V(\phi) \sim \phi^n$. These inflationary model parameters can be measured through  CMB observable associated with curvature, ${\mathcal{R}}$ power spectrum $\Delta^2_{\mathcal{R}}  =A_{\mathcal{R}} (k/k_0)^{n_s -1}$. The amplitude of the spectrum is measured as $A_{\mathcal{R}} = (2.19 \pm 0.06 ) \times 10^{-9}$ normalized at the pivot scale $k_0=0.05 {\rm Mpc}^{-1}$. Another CMB observable is the scale-invariant tensor power spectrum, $\Delta^2_{\mathcal{T}}  =A_{\mathcal{T}}$, with the upper limit on its amplitude, $A_{\mathcal{T}} = r A_{\mathcal{R}} \leq 0.036 \times A_{\mathcal{R}}$. Therefore, we have two inflationary observables $(n_s,r)$, that are expressed in terms of inflaton field through its slow roll parameters $\epsilon$ and $\eta$ as:
\be \label{inflationp}
 n_{s}= 1- 6 \epsilon(\phi)+ 2 \eta(\phi)~,~r=16\epsilon(\phi)\,,
\ee
In any inflation model, the above two CMB observable parameters are typically mapped to two important inflationary quantities, and those are the inflationary energy or Hubble parameter $H_k$, and the inflationary e-folding number $N_k$. Under the slow roll approximation, all are defined at the pivot scale $k=k_0$ as, 
\begin{eqnarray}\label{Hk}
H_{\rm k} &=& \frac{\pi M_{\rm p}\sqrt{r\,A_{\mathcal R}}} {\sqrt{2}} \simeq  \sqrt{\frac{V(\phi_{\rm k})}{3 M_{\rm p}^2} } \\
&\simeq& 2 \times 10^{-5} M_{\rm p} \left[\frac{r}{0.036}\right]^{\frac12}\left[\frac{A_{\rm \mathcal R}}{2.19\times10^{-9}}\right]^{\frac12} , \nonumber \\ 
N_{\rm k}&  =& \int_{\phi_{\rm k}}^{\phi_{\rm end}} \frac{|d\phi|}{\sqrt{2\epsilon(\phi)} M_{\rm p}} ~~.
\end{eqnarray} 
Where, $M_{\rm p} = 2.43 \times 10^{18}$ GeV is the reduced Planck mass. From the above expressions, one can obtain the maximum limit on the inflaton potential in the large field limit, $V(\phi_{\rm k})\simeq \Lambda^4 \rightarrow \Lambda \lesssim 1.4 \times 10^{16}$ GeV. However, more general expressions for $\Lambda$ are available (for $\alpha$ attractor E and T-model see, Ref.\cite{Drewes:2017fmn}, \cite{Garcia:2020wiy} for details) which are corrected by other inflationary parameters $(\alpha, n)$. 
Where $(\phi_{\rm k},\phi_{\rm end})$ represents the inflaton field value at the beginning, which is usually set at the pivot scale and at the end of inflation respectively. The condition of inflation end is defined as $\epsilon(\phi_{\rm end}) = 1$.
More general expressions for the $N_{\rm k}$ and $H_{\rm k}$ in terms of inflationary parameters can be obtained in terms of inflationary parameters (see \cite{Drewes:2017fmn},\cite{Garcia:2020wiy} for details). 
The tensor-to-scalar ratio $r$ turns out as (see, last expression of Eq.\ref{inflationp})
\be
\label{rattrac}
r\;=\;
\begin{cases}
\frac{16n^2}{3\alpha}\left(e^{\sqrt{\frac{2}{3\,\alpha}}\frac{\phi_k}{M_{\rm p}}}-1\right)^{-2}\,,& {\rm E- model},\\[10pt]
\frac{16n^2}{3\alpha} \text{csch}^2\left(\sqrt{\frac{2}{3\,\alpha}}\frac{ \phi _k}{M_{\rm p}}\right)\,,& {\rm T- model}.
\end{cases}
\ee
For the minimal model, $r$ can be calculated numerically. 
Post inflationary dynamics, which we call reheating, will be controlled by the energy of the inflaton ($\rho_{\phi}^{\rm end}$), setting all the other energy components to be negligible,
\be\label{ic}
\rho^{\rm end}_{\phi} 
\sim \frac{\Lambda^4}{\alpha_1^n} \left[\frac{\phi_{\rm end}}{M_{\rm p}}\right]^n = \frac{4.1 \times 10^{64}}{\alpha_1^n} \left[\frac{\phi_{\rm end}}{M_p}\right]^n \left[\frac{\Lambda}{5.8\times10^{-3}M_{\rm p}}\right]^4 ,
\ee
which is in unit of ${\rm GeV}^4$ and  $\alpha_1=(\sqrt{{3\alpha}/{2}}, \sqrt{6\alpha}, \phi_*)$ for $\alpha$-attractor E, T and minimal model respectively. This immediately suggests the maximum value of the Hubble parameter at the end of reheating $H_{\rm end}^{\rm max} \simeq \pi M_{\rm p} \sqrt{r A_{\rm \mathcal{R}}/2} \sim 5\times 10^{13}$ GeV. Any perturbation generated during inflation will evolve through the subsequent phases and may acquire distinct signatures of those phases. We particularly study the reheating phase, which is directly involved with inflaton decay. 
\section{\textbf{Reheating dynamics and constraints}}\label{sc3}
Once inflation ends, the inflaton field coherently oscillates around minimum with the potential $V(\phi) \sim \phi^n$. The coherently oscillating inflaton can be mode decomposed into, $
\phi(t)=\phi_0(t).\mathcal{P}(t)$, with $\phi_0(t)$ representing the decaying amplitude of the oscillation and $\mathcal{P}(t) = \sum_\nu \mathcal{P}_\nu e^{i \nu \Omega t} $ encoding the oscillation of the inflaton with the fundamental frequency calculated to be \cite{Garcia:2020wiy},
\be \label{fre}
\Omega = m_\phi(t) \xi = m_{\phi}(t)\sqrt{\frac{\pi n}{2(n-1)}}\frac{\Gamma\left(\frac{1}{2}+\frac{1}{n}\right)}{\Gamma\left(\frac{1}{n}\right)}\,.
\ee
The new symbol $\xi$ is introduced for later purposes. The effective mass of the inflaton is defined as $m_{\phi}^2 ={\partial^2 V(\phi)}/{\partial \phi^2}|_{\phi_0}$, and we have
\begin{eqnarray} \label{mphi}
m_\phi^2\simeq \frac{(n^2-n)\Lambda^4}{\alpha_1^n M_{\rm p}^2}\left[\frac{\phi_0}{M_{\rm p}}\right]^{n-2} 
\sim {(m_\phi^{\rm end})}^2\left[\frac{a}{a_{\rm end}}\right]^{-6\,w_\phi}\,,
\end{eqnarray}
where $m_\phi^{\rm end}$ is the inflaton mass defined at the end of the inflation
\be\label{massend}
m_\phi^{\rm end}\simeq  \frac{\sqrt{n\,\left(n-1\right)}}{\alpha_1}\frac{\Lambda^{\frac{4}{n}}}{M_{\rm p}}\left(\rho_\phi^{\rm end}\right)^{\frac{n-2}{2\,n}}\,.
\ee
As an example, for $n=6$ and setting $\alpha_1 \sim {\cal O}(1)$, one can obtain,
\be
m_\phi^{\rm end}\sim 4.5\times 10^{14}\,\left[\frac{\Lambda}{1.4\times10^{16}}\right]^{\frac{4}{3}}\,\left[\frac{\rho_\phi^{\rm end}}{4.1\times10^{64}}\right]^{\frac{2}{3}}\,,
\ee
measured in GeV. Radiation fields coupled with inflaton will be produced during this period quantum mechanically, which is called reheating. Associated with this oscillating field, we identify the effective inflaton equation state $w_{\phi} = {(n-2)}/{(n+2)}$. We assume $n \geq 2$. Therefore, during reheating, we will focus on the equation of state within $0 \leq w_{\phi} \leq 1$. The oscillating average energy density of the inflaton is defined with respect to $\phi_0$ as $\rho_{\phi} = (1/2)\langle({\dot \phi}^2 + V(\phi))\rangle = V(\phi)|_{\phi_0}$. In order to solve reheating dynamics, the Boltzmann equations for the energy density of radiation ($\rho_R)$ and inflaton $(\rho_{\phi})$ supplemented with the Hubble equation are,
\begin{eqnarray}
&& \dot{\rho_{\rm \phi}}+3H(1+w_\phi)\rho_\phi= -\Gamma_\phi\rho_{\rm \phi}\,(1+w_{\rm \phi}) ,  \label{Boltzman1} \nno \\
&&\dot{\rho}_{\rm R}+4H\rho_{\rm R}=\Gamma_{\rm \phi}\rho_{\rm \phi}(1+w_{\rm \phi})\,,
\label{Boltzman2} \nno \\
&&H^2=\frac{\rho_{\rm \phi}+\rho_{\rm R}}{3\,M_{\rm p}^2}\label{Boltzman3}\,,
\end{eqnarray}
where, $\Gamma_{\phi}$ is the inflaton decay rate. As stated earlier, inflaton is decaying into radiation, and for our study we consider radiation to be either massless scalar or fermion. We chose two phenomenological decay processes governed by the following interaction Lagrangian,
\be
\mathcal{L}_{\rm int} \supset  \begin{cases} 
h \phi \bar{f}f & \phi \to \bar{f}f \\
 g \phi b^2 & \phi \to b b\\
\end{cases}
\label{RH:procs}
\ee
\begin{figure}[t]
\centering
\includegraphics[width=\columnwidth]{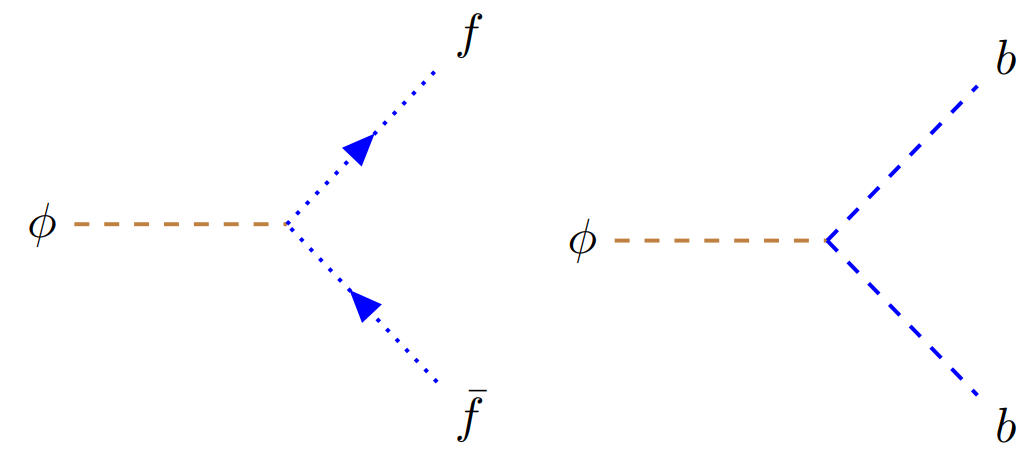}
\caption{\em \small Feynman diagrams for two different inflaton decay channel :(i) $\phi\rightarrow\bar ff$, (ii) $\phi\rightarrow bb$ }
\label{Feynmann}
\end{figure}
with $f\,(b$) standing for fermionic(bosonic) particle. $h$ is the dimensionless Yukawa coupling, and $g$ is a dimensionful bosonic coupling. The associated Feynman diagrams are depicted in Fig.\ref{Feynmann}. Note that, although our analysis will be limited to these two scenarios, using the same methodology, one can analyze other inflaton-radiation couplings. 
Incorporating the inflaton oscillation effect, the effective inflaton decay rate $\Gamma_{\phi}$ associated with these two processes have been computed \cite{Garcia:2020wiy,Haque:2023yra}, and can  be expressed in terms of inflaton energy density as
\be
\label{Eq:gammaphibis}
\Gamma_{\phi} = 
\begin{cases}
\sqrt{\frac{2n(n-1)}{3\alpha}}  \left(\frac{\Lambda}{M_p}\right)^{\frac {4}{n}} \dfrac{M_p h_{\rm eff}^2}{8\pi} \left(\frac{\rho_{\phi}}{M_{\rm p}^4}\right)^{\frac{n-2}{2n}}, ~& \phi\rightarrow \bar{f}f\\
\sqrt{\frac{3\alpha}{2n(n-1)}} \left(\frac{M_p}{\Lambda}\right)^{\frac {4}{n}} \dfrac{g_{\rm eff}^2}{8\pi M_p} \left(\frac{\rho_{\phi}}{M_{\rm p}^4}\right)^{\frac{2-n}{2n}},~& \phi\rightarrow bb,
\end{cases}
\ee
The ratio of the oscillation induced effective coupling parameters $h_{\rm eff}$ and $g_{\rm eff}$ over their respective tree level values are calculated to be \cite{Garcia:2020wiy}
\begin{eqnarray}{\label{t2}}
&&\left(\frac{g_{\rm eff}}{g}\right)^2=(n+2)(n-1)\,\xi\sum^{\infty}_{\nu=1}\nu\lvert\mathcal P_\nu\rvert^2\\
&&\left(\frac{h_{\rm eff}}{h}\right)^2=(n+2)(n-1)\,\xi^3\sum^{\infty}_{\nu=1}\nu^3\lvert\mathcal P_\nu\rvert^2.
\end{eqnarray}
\begin{table}[t]
\scriptsize{
\caption{Numerical values of the Fourier sums in the effective couplings:}\label{fouriersum}
\centering
 \begin{tabular}{||c | c |c |c |c||} 
 \hline
 $n(w_\phi)$ & $\sum \nu\lvert\mathcal P_\nu\rvert^2$ & $\sum \nu^3\lvert \mathcal{P}_\nu\rvert^2$ & $\frac{g_{\rm eff}}{g}$ & $\frac{h_{\rm eff}}{h}$\\ [0.5ex] 
 \hline\hline
 2 (0.0) & $\frac{1}{4}$ & $\frac{1}{4}$ & 1 & 1\\ 
 4 (1/3) & 0.229  & 0.241 & 1.42 & 0.71\\
 10 (0.67)& 0.205 & 0.256 & 2.13 & 0.49 \\
 20 (0.82) & 0.191 & 0.286 & 2.92 & 0.38\\
 400 (0.99) & 0.174  & 0.358 &  12.5 & 0.10 \\ [1ex] 
 \hline
 \end{tabular}}
\end{table}
In the table-\ref{fouriersum}, we have tabulated the effective coupling constant with respect to tree label one. It is interesting to observe that the oscillation with a higher $n$ value effectively enhances the bosonic production rate but significantly diminishes the fermionic production rate. For example, going from $n=2$(matter like) to $n=10$ ($w_\phi=0.67$) for inflaton, the tree-level bosonic coupling gets doubled $g_{\rm eff} \simeq  2 g$, whereas that of the fermionic coupling reduced by half $h_{\rm eff} \simeq 0.5 h$.    
With all these ingredients, we have solved  the coupled Boltzmann Eq.\ref{Boltzman3} numerically with the appropriate boundary condition. For analytical estimation, we also obtain the approximate analytical solution for different energy density components as
\begin{eqnarray}\label{rhophi}
&&\rho_\phi \simeq 
\rho^{end}_{\phi}\left(\frac{a}{a_{end}}\right)^{-3(1+w_{\phi})} ,\\
&&\rho_{\rm R} \simeq
\begin{cases}
\frac{\rho_\phi^{\rm end}(1+w_\phi)m^{\rm end}_\phi h_{\rm eff}^2}{4\pi(5-9w_\phi)\,H_{\rm end}}  \left[\left(\frac{a}{a_{\rm end}}\right)^{\frac{-3-9w_\phi}{2}}-\left(\frac{a}{a_{\rm end}}\right)^{-4}\right]~\,\nonumber\\
\frac{\rho_\phi^{\rm end}(1+w_\phi) g_{\rm eff}^2}{4\pi(5+3w_\phi) m^{\rm end}_\phi H_{\rm end}} \left[\left(\frac{a}{a_{\rm end}}\right)^{\frac{-3+3w_\phi}{2}}-\left(\frac{a}{a_{\rm end}}\right)^{-4}\right] .
\end{cases}
\end{eqnarray}
Here, $(H_{\rm end}, m_{\phi}^{\rm end})$ are the Hubble constant, inflaton mass calculated at the end of the inflation.
With this solutions, we can identify important physical quantities, namely, reheating temperature $T_{\rm re}$, defined at the end of reheating at $\rho_{\phi} = \rho_R \propto T_{\rm re}^4$. At the reheating point, the energy density of radiation will be
\begin{eqnarray}
   T_{\rm re} &\simeq& \left(\frac{30 \rho_\phi^{\rm end}}{\pi^2 g_*^{\rm re}}\right)^{\frac 14}\,e^{-\frac 3 4 N_{re}\,(1+w_\phi)} \\ 
&=& 6\times10^{15} e^{-\frac {3N_{\rm re}} {4} (1+w_\phi)} \left[\frac{\rho_{\phi}^{\rm end}}{4.1\times10^{64}}\right]^{\frac14} 
\end{eqnarray}
Where, $g_*^{\rm re}$ is the effective number of relativistic degrees of freedom at the end of reheating, and we take $g_*^{\rm re} =100$ though out the paper. From the above equation, it is clear that we have an upper limit on the reheating temperature $T_{\rm re}^{\rm max} \sim 10^{15}$ GeV, for $N_{\rm re} =0$, dubbed as instantaneous reheating, and as expected that does not depend on the $w_{\phi}$. As mentioned earlier in the introduction, there exists a naive lower limit on the temperature $T^{\rm min}_{\rm re} = T_{\rm BBN} = 4 {\rm MeV}$ \cite{Kawasaki:2000en,Hannestad:2004px}, which can immediately give us the relation between the maximum reheating e-folding number associated with BBN constraint as,
\be
    N_{\rm re}^{\rm max} \simeq\frac{4}{3(1+w_\phi)}\ln \left[2.5\times 10^{18} \left[\frac{\rho_{\phi}^{\rm end}}{4.1\times10^{64}}\right]^{\frac14} \right]\,.
\ee
This clearly suggests that the maximum reheating e-folding number decreases with the equation of state. Now as an example, for $\rho_\phi^{\rm end}\sim 4.1\times 10^{64}$ $\rm GeV^4$, $N_{\rm re}^{\rm max}$ simply turns out as $\sim {56.5}/{(1+w_\phi)}$. $N_{\rm re}^{\rm max} \sim 56.5$ for a matter like inflaton equation of state ($w_{\phi} =0$) and it reduces to half $N_{\rm re}^{\rm max} \sim 28.2$ for kination like state ($w_{\phi} =1$).

Post-reheating history is also important in order to put constrain on the inflationary e-folding number.
The generic assumption after the reheating phase is that the comoving entropy density remains conserved, and such conservation law starting from reheating end to the present day imposes an additional relation among the parameters ($N_k,\,N_{\rm re},\,T_{\rm re}$) as follows \cite{Dai:2014jja,Cook:2015vqa}, 
\be \label{entropy-conservation}
T_{\rm re}=\left(\frac{43}{11\,g_*^{\rm re}}\right)^{1/3}\,\left(\frac{a_0\,H_{ k}}{k}\right)\,e^{-(N_k+N_{\rm re})}\,T_0\,,
\ee
where the present CMB temperature is $T_0=2.725$ K. Considering $H_k\sim H_{\rm end}$ (which may not true for higher values of $\alpha(\phi_*)$), expression for $N_{\rm re}$ can be expressed in more simplified manner,
\be
N_{\rm k}=\ln \left[1.2\times 10^{40}\left[\frac{\rho_{\phi}^{\rm end}}{4.1\times10^{64}}\right]^{\frac12}\frac{1}{T_{\rm re}}\right]-N_{\rm re}\,.
\ee
Here $T_{\rm re}$ is measured in GeV unit. From the above relation, if we consider $\rho_\phi^{\rm end}\sim 4.1\times 10^{64}$ $\rm GeV^4$ and $N_{\rm re} \sim 0$ (instant reheating) that immediately sets the maximum probable reheating temperature $T^{\rm max}_{\rm re} \sim 10^{15}$ GeV and the inflationary e-folding number for the CMB pivot scale as $N_{\rm k} \sim 57.8$. The most important point of the relation Eq.\ref{entropy-conservation} probably is that for a given model, it offers an interesting relation between inflaton-radiation coupling following the relation $T_{\rm re} \propto \sqrt{\Gamma_{\phi}}$ with the inflationary observable $n_s$ through the e-folding number $N_{\rm k}$, which is apparent from the Figs.\ref{nsrE}-\ref{w13}. For the special case, $w_\phi=1/3$, $N_{\rm k}$ turns out to be independent of the reheating temperature $T_{\rm re}$ (see Fig.\ref{w13}). For a fixed inflation model $N_k$ solely depends on inflation parameter as follows, \cite{Cook:2015vqa,Haque:2020zco}
\be
N_{\rm k}=40.26-\ln\left(\frac{{H_{\rm end}}^{1/2}}{H_{\rm k}} \right)\,.
\ee
The above equation implies that $N_{\rm k}$ is only sensitive to the $H_k$ and $H_{\rm end}$ value, which is regulated by the potential parameter $\alpha (\phi_*)$. The  limiting value of $\alpha (\phi_*)$, $H_{\rm end}$ and $N_{\rm k}$ are shown in Table \ref{infresE0}.\\
So far, our discussions are on two main points, the nature of large-scale inflationary fluctuations imprinted into ($n_s,r)$ and the background reheating dynamics constraining the reheating parameters.  
However, the dynamics of high-frequency modes, particularly of PGWs, turn out to offer stronger bound on lower reheating temperatures than $T_{\rm BBN}$, particularly for a stiff equation of state $w_{\phi}>0.6$.   
 \section{PGWs and constraints}\label{sc4}
 Because of its naturally weak coupling, GWs prove to be an outstanding probe to look into the early universe, which can be as early as inflation and reheating. We wish to probe the reheating phase particularly and see how PGWs play out, yielding stronger constraints on the inflaton-radiation coupling by raising the lower limit of reheating temperature. However, to probe the reheating phase's typical frequency range, one requires a wide span of $10^{11} > f > 10^{-10}$ Hz, which is way outside the window of the CMB spectrum. Here, the frequency, $f$ associated with a particular mode $k$ is related by $f=2\pi/k$. Therefore, along with the constraints on the large-scale fluctuation by PLANCK, BICEP/$Keck$, we consider high-frequency PGWs  to obtain further constraints on the inflaton model. Interesting to note that a large number of proposed/ongoing GW detectors are designed within this wide range of frequencies \cite{LIGOScientific:2016jlg,Punturo:2010zz,Crowder:2005nr,Seto:2001qf,LISA:2017pwj,Janssen:2014dka}, and there is a growing anticipation that the stochastic GW background may provide us hints about the physics operating in the early universe, including reheating. PGWs generated from the quantum vacuum during inflation evolve through the different phases of the universe, including the epoch of reheating, until we observe them today~\cite{Mishra:2021wkm,Haque:2021dha,Vagnozzi:2020gtf,Benetti:2021uea}. The amplitude and the evolution of the PGWs spectrum are sensitive to the energy scale of the inflation and the post-inflationary reheating equation of state $w_\phi$.
For those modes between $k_{\rm re} < k < k_{\rm f}$ which become sub-Hubble at some time during reheating, the PGW spectrum at the present time assumes the following form, 
(see \cite{Haque:2021dha} for detailed calculation)
\be \label{GW}
\Omega^{\rm k}_{\rm GW}h^2\simeq \Omega^{\rm inf}_{\rm GW}h^2 \frac{\mu (w_\phi)}{\pi}\left(\frac{k}{ k_{\rm re}}\right)^{-\frac{(2-6\,w_\phi)}{(1+3\,w_\phi)}}
\ee
Where, $\mu(w_\phi)=(1+3\omega_\phi)^{\frac{4}{1+3\,w_\phi}}\,\Gamma^2\left(\frac{5+ 3\,w_{\phi}}{2+6\,w_{\phi}} \right)$, which typically ${\cal O}(1)$ value.
Therefore, the above equation contains two main components. The scale-invariant part,
 controlled by the inflationary energy scale, 
\be
\Omega^{\rm inf}_{\rm GW}h^2
= \frac{\Omega_{\rm R} h^2 H_{\rm end}^2}{12 \pi^2 M_{\rm p}^2}\, = 3.5 \times 10^{-17} \left(\frac{H_{\rm end}}{10^{-5} M_p}\right)^2.
\ee
Where we used the present radiation abundance $\Omega_{\rm R} h^2= 4.16\times10^{-5}$, the second part is the scale-dependent one, which encodes crucial information about reheating. $k_{\rm re}$ is the mode that enters the horizon at the end of reheating, and its value naturally depends on $T_{\rm re}$. On the other hand, $k_{\rm f}$ enters the reheating phase at the beginning and is fixed by $H_{\rm end}$. We can clearly see from Eq.\ref{GW} that modes entering during reheating becomes red tilted for $w_{\phi} <1/3$, and blue tilted for $w_\phi>1/3$. Thus, amplitude of the GW energy $\Omega^{\rm k}_{\rm GW}h^2$ is an increasing function of $k$ for $w_\phi>1/3$.
In the next part, we will discuss the calculation part of the BBN constraints in the context of PGWs

 The effective number of additional relativistic degrees of freedom quantified by $\Delta N_{\rm eff}$ at the time of BBN place tighter constraints on the reheating temperature. High-frequency GWs can be thought of as an effective relativistic degree of freedom during reheating. Therefore, if the reheating phase is long enough to correspond to low reheating temperature, $k_{\rm f}>>k_{re}$, the GW energy puts a tighter constraint. 
The constraint equation is expressed as,
\begin{align}
\Delta N_{\rm eff} \geq \frac {1}{\Omega_{\rm R}h^2} \frac{8}{7}\left(\frac{11}{4}\right)^{4/3} \int_{k_{\rm 0}}^{k_{\rm f}}\frac{dk}{k}\Omega_{\rm GW}^{\rm k}\,h^2(k)
\end{align}
\begin{figure}[t]
\centering
\includegraphics[width=\columnwidth]{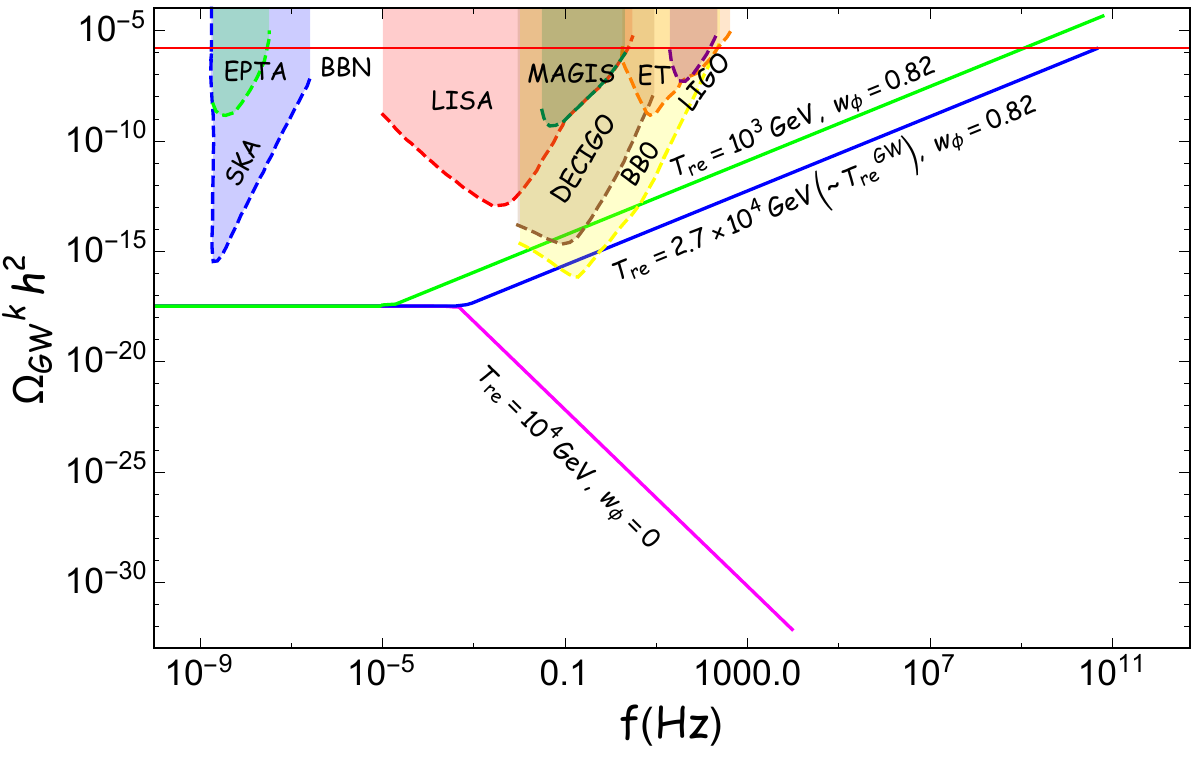}
\caption{\em \small Variation of the dimensionless energy density of PGWs observed today, viz $\Omega_{GW}^{\rm k}$ as a function of frequency over a wide range for $\alpha$-attractor E model ($\alpha=1$). }
\label{GW1}
\end{figure}
The combination of the latest \textcolor{red}{Planck} 2018 + BAO data provides $\Delta N_{\rm eff}\leq 0.284$ (within 
$2$-$\sigma$)~\cite{Planck:2018vyg}. 
The straight forward's calculation with the above equation yields the following bound on the total GW energy density  $\Omega_{\rm GW}h^2 < 1.7\times 10^{-6}$, and that is depicted by the solid red line in Fig.\ref{GW1}. However, we wish to translate this bound into the associated reheating temperature. 
In the above expression, it is the maximum $k \sim k_{\rm f}$ which contributes the most to the total GW energy, and one obtains the following constraint relation,
\be \label{bbncon}
2.1\times 10^{-11} \frac{\mu (w_\phi) (1+3\,w_\phi)} {\pi(6\,w_\phi -2)}  \left[\frac{H_{\rm end}}{10^{-5} M_{\rm p}}\right]^2 \leq \left(\frac{k_{\rm f}}{ k_{\rm re}}\right)^{\frac{(2-6\,w_\phi)}{(1+3\,w_\phi)}}
\ee
The relation between the two scales $(k_{\rm f}, k_{\rm re}$) can be further expressed in terms of reheating temperature as follows: 
\be
\left(\frac{k_{\rm f}}{k_{\rm re}}\right)=\left(\frac{15\,\rho_\phi^{\rm end}}{\pi^2\,g_*^{\rm re}}\right)^{\frac{1+3 w_\phi}{6\,(1+w_\phi)}}\,T_{\rm re}^{\frac{-2 (1+3 w_\phi)}{3(1+w_\phi)}}\,.
\ee
 Finally, using the above relations into Eq.\ref{bbncon}, we obtain a lower limit on the reheating temperature, particularly when the inflaton equation of state $w_{\phi} >1/3$, 
\begin{eqnarray}
T_{\rm re}&>&0.35 \left(\frac{45.6\,M_p^4}{\mu(\phi)}\right)^{\frac{3\,(1+w_\phi)}{4(1-3\,w_\phi)}}\left(\rho_\phi^{\rm end}\right)^{-\frac{1}{2}\,\frac{(1+3\,w_\phi)}{(1-3\,w_\phi)}}\nonumber\\
&=& T_{\rm re}^{\rm GW}.
\end{eqnarray}
Setting the above temperature with BBN energy scale $T_{\rm re}^{\rm GW}\sim T_{\rm BBN} (4)$ MeV, we can see that the BBN bound of PGWs only important when $w_\phi\geq0.60$. We symbolize this new lower limit on reheating temperature from PGW as $T_{\rm re}^{\rm GW}$. As an example for $w_\phi=0.82$ ($n=20$) the expression for $T_{\rm re}^{\rm GW}$ can be expressed as 
\be
T_{\rm re}^{\rm GW}\sim 2.1\times 10^6\,\left(\frac{\rho_\phi^{\rm end}}{4.1\times 10^{64}}\right)^{\frac{19}{16}}\,.
\ee
\begin{table}[t]
\scriptsize{
\caption{Numerical values of the $T_{\rm re}^{\rm GW}$ for a fixed value of $H_{\rm end}=10^{13}$ GeV:}\label{GWtre}
\centering
 \begin{tabular}{||c | c ||} 
 \hline
 $n(w_\phi)$ & $T_{\rm re}^{\rm GW}$ (GeV) \\ [0.5ex] 
 \hline\hline
 8 (0.60) & $1.8\times 10^{-2}$ \\
  10 (0.67) & $13.4$ \\
   12 (0.71) & $3.6\times 10^2$ \\
 20 (0.82) & $5.0\times 10^4$ \\ 
  50 (0.92) & $1.2\times 10^6$ \\
   100 (0.98) & $4.7\times 10^6$ \\
     400 (0.99) & $5.7\times 10^6$ \\[1ex] 
 \hline
 \end{tabular}}
\end{table}
Now for $H_{\rm end}=10^{13}$ GeV, $T_{\rm re}^{\rm GW}$ simply turns out as $5\times 10^4$ GeV, and that will set the lower limit on the reheating temperature.
In Fig.\ref{GW1}, we showed blue tilted behavior for $w_\phi=0.82$, for two different sample reheating temperature $(10^4, 10^3)$ GeV. For $T_{\rm re}=10^3$ GeV, $\Omega^{\rm k}_{\rm GW}h^2 \sim 10^{-4}$ at $k=k_{\rm f}$, and it clearly violates the BBN constraints discussed just above. However, for $T_{\rm re} \sim 2.7\times 10^4$ GeV, we see $\Omega^{{\rm k}_{\rm f}}_{\rm GW}h^2$ marginally satisfies the BBN constraints, and that is consistent with our above estimation of $T_{\rm re}^{\rm GW}$.

 Fixing the same $H_{\rm end}$, the numerical values of $T_{\rm re}^{\rm GW}$ for different EoS shown in Table-\ref{GWtre}. Since $T_{\rm re}^{\rm GW}$ is a function of both $H_{\rm end}$ and $w_\phi$, for a fixed value of $w_\phi$ it can restrict inflationary energy scale, which in turn puts a constraint on the potential parameter such as $\alpha$ and $\phi_*$ for the attractor and minimal model respectively. Another interesting point is that this $T_{\rm re}^{\rm GW}$ also set bounds on the inflaton coupling to other fields, which we will discuss in more detail in the subsequent sections for $\phi\to\bar{f}{f}/bb$ decay processes.
\section{One-loop effective potential and perturbative constraints during inflation}\label{sc5}
In the previous section, we discuss about the possible lower bound on $T_{\rm re}$ though PGWs. Purely from the perturbative reheating point of view, the existence of maximum $T_{\rm re}$ is typically attributed to the instant reheating process with $N_{\rm re} \to 0$. In this regard, one should remember that the upper limit on $T_{\rm re}$ is directly related to the upper limit on inflaton-radiation coupling. During inflation, such coupling can naturally modify the effective inflaton potential through loop correction, which may modify the aforesaid upper bound so that inflation is not disturbed. In this section, we investigate the upper bound of the coupling parameters, below which the inflationary scenario is not affected by the inflaton-radiation coupling. We follow Coleman and Weinberg's (CW) 1-loop radiative correction formalism to determine the bound. The 1-loop corrected inflaton-potential is given by \cite{Coleman:1973jx,Drees:2021wgd}
\begin{equation}
    V_{\rm CW}(\phi)=\sum_{i}\frac{n_i}{64\pi^2}(-1)^{2s_i}m^4_i(\phi)\left[\ln\left(\frac{m^2_i(\phi)}{\mu^2}\right)-3/2\right],
\end{equation}
\begin{figure}[t]
\centering
\includegraphics[width=\columnwidth]{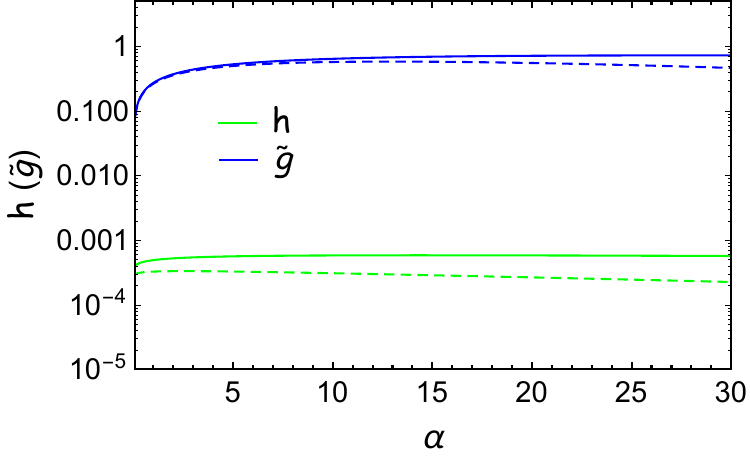}
\caption{\em \small Here we plot the variation of the upper limit of the coupling parameter (from Eq.\ref{critgh}) as a function  $\alpha$ for $w_\phi=0.0$ (solid) and $w_\phi=0.99$ (dashed) for $\alpha$-attractor E model.}
\label{VCW}
\end{figure}
where summation, $i$ runs over the radiation  fields ($f,b$). $(n_i,s_i)$ represents the total internal degrees of freedom and spin.  $\mu$ is the renormalization scale which we have taken $\phi_k$. $m_i$ corresponds to   inflaton field induced mass. 
The field-dependent mass of the fermionic (f) and the  bosonic (b) fields can be written as 
\begin{equation}
  m^2_i(\phi) = \begin{cases} 
h^2 \phi^2 \,& ~~{\rm for}~~\phi \bar{f}f \\
2\,g \phi &~~{\rm for}~~\phi b b \\
\end{cases}
\end{equation}
 \begin{figure}[t]
         \begin{center}
\includegraphics[width=8.65cm,height=3.1cm]
          {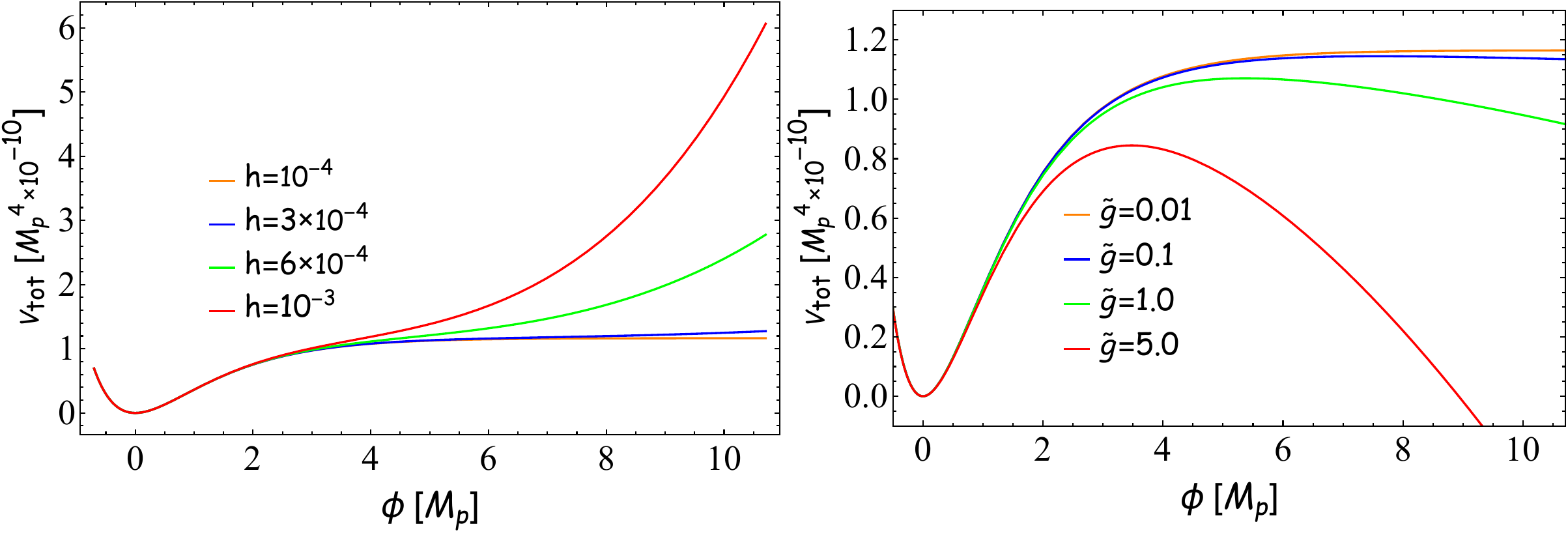}
          \caption{\em Sum of the tree level potential and CW 1 loop corrected potential, $V_{\rm tot}=V_{\rm tree}+V_{\rm CW}$ as a function of filed value $\phi$ for $\alpha$-attractor E-model ($\alpha=1,\,n=2$) with two different interactions $h\phi\bar ff$ (left panel) and $g\phi bb$ (right panel). }
          \label{couplingcw}
          \end{center}
      \end{figure}
For the stability of the inflation potential, the coupling parameter ($g,h$) should be such that the following inequality must hold,

\begin{equation}{\label{stacr}}
   \big | V_{\rm tree}(\phi) \big |>\big|V_{\rm CW}(\phi)\big| ,
\end{equation}
where, $V_{\rm tree}$ is the tree level potential defined in Eq.\ref{attractorpotential}. Utilizing the above condition (Eq. \ref{stacr}), one can find the following analytical expressions for the upper bound of the coupling parameters are
\begin{eqnarray}{\label{critgh}}
  && g <\frac{8i\sqrt{2}\pi\lvert V_{\rm tree}(\phi)\rvert^{1/2}}{\phi^2\mathcal{W}_{-1}[-\frac{128\pi^2\phi^2}{e^3\mu^4}\lvert V_{\rm tree}(\phi)\rvert]^{1/2}}  \nno \\
       && h < \frac{(-2)^{1/4}\sqrt{\pi}\lvert V_{\rm tree}(\phi)\rvert^{1/4}}{\phi\mathcal{W}_{-1}[-\frac{32\pi^2}{e^3\mu^4}\lvert V_{\rm tree}(\phi)\rvert]^{1/4}}
\end{eqnarray}
where $\mathcal{W}_{-1}$ is the Lambert function of branch $-1$ and the above expressions are only for E and T model, so $V_{\rm tree}$ is different for different models. For the minimal model, we could not obtain such an analytical expression. The Eq.\ref{critgh} seems to suggest a complicated dependence of the couplings on the inflaton parameters. However, Fig-\ref{VCW} indicates that the lower limit on the couplings is indeed insensitive to $\alpha$ or $w_\phi$, and approximately we, therefore, estimate,
\begin{equation}{\label{critgh}}
   \begin{aligned}
     &g<8\times 10^{13}\,\left[\frac{\Lambda}{1.4\times10^{16}}\right]^2\,\left[\frac{M_{\rm p}}{\phi}\right]\,,\\
    &h<5.8\times 10^{-3}\,\left[\frac{\Lambda}{1.4\times10^{16}}\right]\,\left[\frac{M_{\rm p}}{\phi}\right]\,,
\end{aligned}
\end{equation}
We have checked numerically that this bound is also the same for the E, T, and minimal models. We define all the inflationary observable at the pivot scale with a field value, $\phi_{\rm k}$. In Fig.\ref{couplingcw}, we have shown how a strong coupling parameter modifies the original potential through CW one loop radiative correction, and violation of the above condition Eq.\ref{critgh} always leads to a deviation from the original potential. Therefore, any coupling violating Eq.\ref{critgh}   will not be allowed. In all the plots, we shaded it in blue color and mentioned  it as not-allowed. For example, 
the upper bound of the dimensionless bosonic coupling parameter $\tilde g=g/{m_\phi^{\rm end}}$ assumes  $\mathcal{O}(10^{-1})$ value, and this bound is nearly model-independent. Corresponding $T_{\rm re}^{\rm max}$ value appears to be the same as that of the instantaneous reheating temperature ($\sim 10^{15}$ GeV). Hence, for bosonic reheating no-parameters space is ruled out by the radiative CW correction. On the other hand, for fermionic reheating, the upper bound on the Yukawa coupling parameter $(h)$ assumes nearly model-independent value $\sim 10^{-4}$, and that can be observed from all the Figs.\ref{Tree},\ref{Tret} and \ref{Trem} (shaded in blue color). For the sake of brevity, we call these as Coleman-Weinberg constraints (CWc).

As discussed above, the CWc constraints are about the perturbative correction to the inflaton potential, particularly during inflation. This limit should be assumed as the strict upper limit of the inflaton coupling constant with the radiation. However, during reheating, the process through which inflaton decays into radiation may be nonperturbative in nature. We now turn to discuss another bound on the coupling parameters coming from the  effect during reheating. 

\section{ constraints during reheating}\label{sc6}
Radiation can be produced resonantly during reheating if the inflaton-radiation coupling is strong enough.  
However, our present analysis is perturbative. Therefore, it is imperative to identify the  constraints (NPc) on the coupling  for which our conclusion may not be strictly valid. For this, one usually solves the radiation field equation in the oscillating inflaton background and identifies the coupling region where broad parametric resonance occurs. The mode equation for the radiation field assumes a general Hill-type equation as follows,
\begin{align}\label{fluctueq}
\frac{1}{(m_{\phi}^{\rm end})^2}\frac{d^2 X_{\rm k}}{dt^2}+{\cal Q}(t)^2 X_{\rm k}=0 
\end{align}
Where, $X_{\rm k}$ is a particular field mode for fermion $(f)$, or boson $(b)$, and the associated time-dependent frequencies are\cite{Kofman:1997yn,Lozanov:2019jxc,Dufaux:2006ee,Greene:2000ew,Greene:1998nh};  
\be \label{resop}
{\cal Q}(t)^2 = \begin{cases} 
 \frac{k^2}{(m_{\phi}^{\rm end})^2 a^2}+ q^2(t) \mathcal{P}^2(t)-i\frac{q(t) \dot{\mathcal{P}}(t)}{m_{\phi}^{\rm end}} & ~~{\rm for}~~\phi \bar{f}f \\
\frac{k^2}{(m_{\phi}^{\rm end})^2 a^2}+ q^2(t) \mathcal{P}(t) &~~{\rm for}~~\phi b b \\
\end{cases}
\ee
Given the inflaton-radiation coupling of our interest Eq.\ref{RH:procs}, the resonance parameter $q$ is identified as 
\be\label{qt}
q(t) = \begin{cases} 
 \sqrt{\frac{h^2 \phi^2_0(t)}{(m_{\phi}^{\rm end})^2}} & ~~{\rm for}~~\phi \bar{f}f \\
\sqrt{\frac{g \phi_0(t)}{(m_{\phi}^{\rm end})^2}} &~~{\rm for}~~\phi b b \\
\end{cases}
\ee
\begin{figure*}[t]  
\includegraphics[width=0014.50cm,height=010.5cm]
          {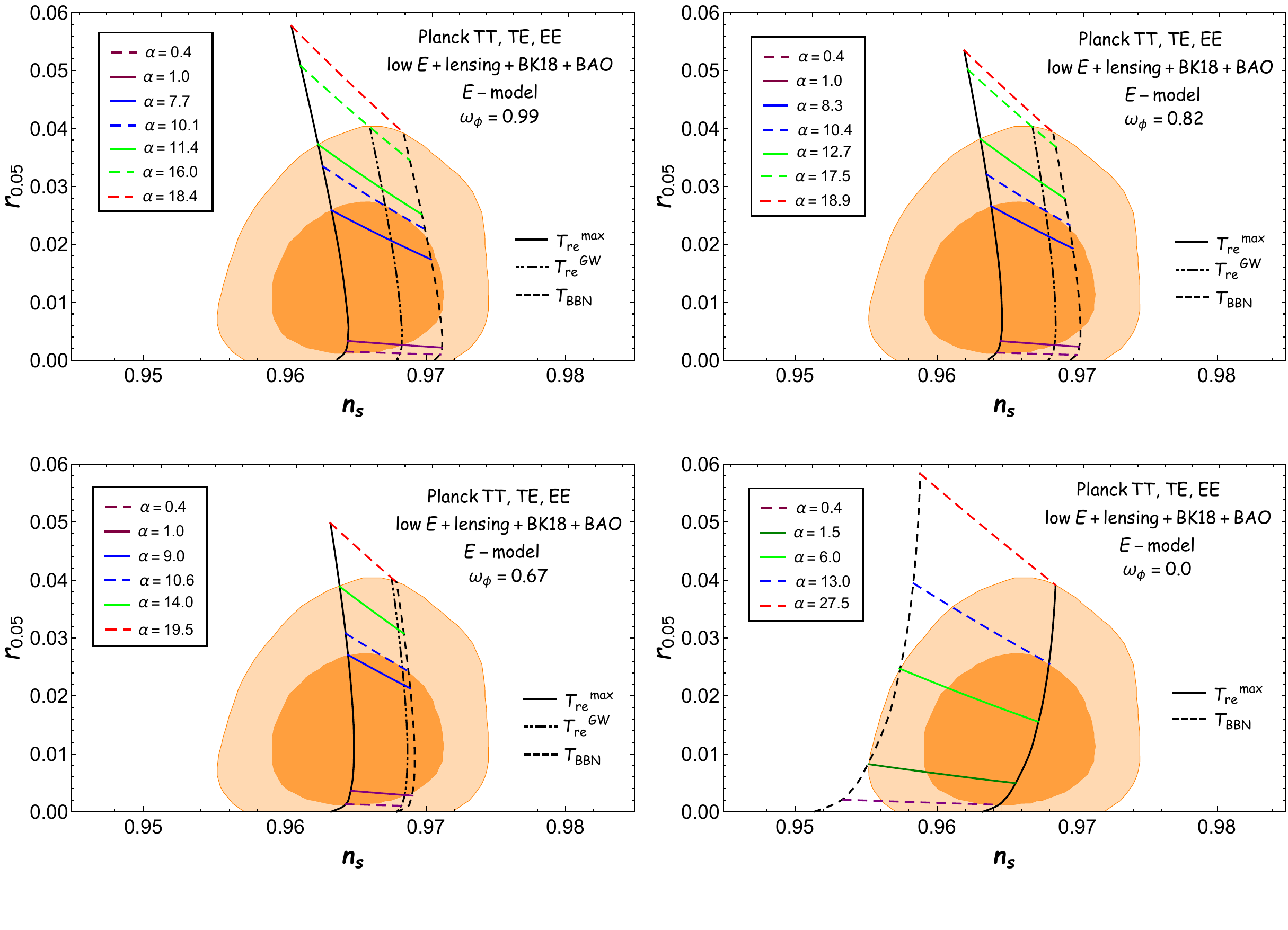}
   \caption{\em Prediction of $\alpha$-attractor E-model  for different $(\alpha,\,w_\phi)$ projected on the recent {\rm PLANCK + BICEP/$Keck$} constrained on $(n_s,r)$ plane Ref.\cite{Tristram:2021tvh}. Deep orange and light orange shaded regions correspond to $1\sigma$ region at 68$\%$ CL and $2\sigma$ region at 95$\%$ C.L., respectively. Reheating temperature varies from $T_{\rm BBN}\to T_{\rm re}^{\rm max}$, showing in solid and dashed black lines. Another important temperature scale, $T_{\rm re}^{\rm GW}$, is shown in a dot-dashed black line.}
\label{nsrE}
\end{figure*}
In the literature, conditions of resonance is usually derived in Minkowski background \cite{Kofman:1997yn,Greene:1997fu,Greene:2000ew}, and the resonance is broadly classified into $q>1$ called broad resonance regime and $q \lesssim 1$ called narrow or no resonance regime. However, in reality, the resonance parameter $q$ depends non-trivially on time through inflaton oscillation amplitude $\phi_0(t)$, and hence naive Minkowski approximation has been observed to be insufficient. Particularly, with increasing reheating equation of state, the inflaton amplitude dilutes very fast, $\phi_0(t)\propto \phi_{\text{end}} ({a(t)}/{a_{\text{end}}})^{-{6}/{(n+2)}}$ depending on different $n$ values. Using decaying inflaton amplitude in $H^2\simeq {V(\phi_0)}/{3M_{p}^2}$, we get the leading order behavior of post-inflationary scale factor as
\begin{equation}\label{scaleforn}
  a(t)=  a_{\text{end}} \left(\frac{ t}{ t_{end}}\right)^{\frac{n+2}{3n}} .
\end{equation}
Using above equation and Eq.\ref{qt} the resonance parameter $q(t)$ evolves as,
\be\label{qtA}
q(t) = \begin{cases} 
 \sqrt{\frac{h^2 \phi^2_{\text{end}}}{(m_{\phi}^{\rm end})^2}}\left(\frac{t}{t_{end}}\right)^{-\frac{2}{n}} & ~~{\rm for}~~\phi \bar{f}f \\
\sqrt{\frac{g \phi_{\text{end}}}{(m_{\phi}^{\rm end})^2}}\left(\frac{t}{t_{end}}\right)^{-\frac{1}{n}} &~~{\rm for}~~\phi b b .
\end{cases}
\ee
In this dynamical scenario, we propose resonance condition as follows:\\ Resonant particle production is strongly related to the violation of adiabaticity condition, and that occurs when the background crosses zero during oscillation. To have significant resonant production within a certain period, one needs to satisfy two essential conditions. The first one is to have the oscillatory background, executing few oscillations within the period of interest. The second condition is that within that period, the resonance $q$-parameter should remain greater than unity. Combining these aforesaid conditions, we state that for broad resonance to take place while the resonance parameter $q$ evolves from its initial value to unity, it must complete at least one oscillation.  Having this dynamical condition of broad resonance, we derive the lower bound of the inflaton-radiation couplings for general background EoS.\\ To compute the number of oscillations required for $q$-parameter to change from its large initial value $q_{\text{in}}$ at the end of inflation to unity,  we measure the dimensionless time-period of the oscillating inflaton as $T^{(\Omega)} = 2\pi m_{\phi}^{\text{end}}/\Omega_0$, where $\Omega_0$ is the background oscillation frequency calculated at $\phi_0 = M_{p}$(see Eq.\ref{fre} and Eq.\ref{mphi} ). On the basis of these, the number of oscillations $N_{\rm osc}$ becomes,
\begin{align}\label{noosci}
     N_{\text{osc}}= \frac{t -t_{end}}{T^{(\Omega)}} =
   \begin{cases}
  & \frac{t_{end}}{T^{(\Omega)}}\left(\left(\frac{h\phi_{\text{end}}}{m_{\phi}^{\text{end}}}\right)^{\frac{n}{2}}-1\right)\\
   &\frac{t_{end}}{T^{(\Omega)}}\left(\left(\frac{\sqrt{g\phi_{\text{end}}}}{m_{\phi}^{\text{end}}}\right)^n-1\right) ,
   \end{cases}
\end{align}
Where
\[\frac{t_{end}}{T^{(\Omega)}}=\frac{\sqrt{n(n-1)}(n+2)M_{\rm p} \Omega_{0}} {2\pi\sqrt{3} nm_{\phi}^{\text{end}}\phi_{\text{end}}}.\] 
Therefore, the minimum criterion that has to be met to achieve efficient resonance is $N_{\text{osc}}>1$. This yields the lower bound of coupling strength for two decay channels as
\begin{eqnarray}\label{lowerboundgh}
h &>& \frac{m_{\phi}^{\rm end}}{\phi_{\text{end}} }\bigg[1+\frac{2\pi\sqrt{3} n\phi_{\text{end}}m_{\phi}^{\rm end}}{\sqrt{n(n-1)}(n+2)M_{\rm p} \Omega_0}\bigg]^{\frac{2}{n}}\nonumber\\
 g &>& \frac{(m_{\phi}^{\rm end})^2}{\phi_{\text{end}} }\bigg[1+\frac{2\pi\sqrt{3}  n\phi_{\text{end}}m_{\phi}^{\rm end}}{\sqrt{n(n-1)}(n+2)M_{\rm p} \Omega_{0}}\bigg]^{\frac{2}{n}}
 \,\label{nonper2}
\end{eqnarray}
The first term on the right-hand side of the above expressions is the one that comes from the standard analysis in Minkowski space. The new bracketed  correction term originates from our dynamical definition. Considering the attractor model, the typical upper limit on  the dimensionless coupling parameter $\tilde{g},h\sim (10^{-4}, 10^{-3}) $ with the allowed range of $w_\phi=(0,\,1)$. To this end, we would like to stress that unlike CWc discussed before, 
NPc only suggests that above this coupling, the reheating dynamics will be modified by the  effect.  

\section{Results and discussions}\label{sc7}

      \begin{figure*}[t!]
         \begin{center}
\includegraphics[width=0014.50cm]
          {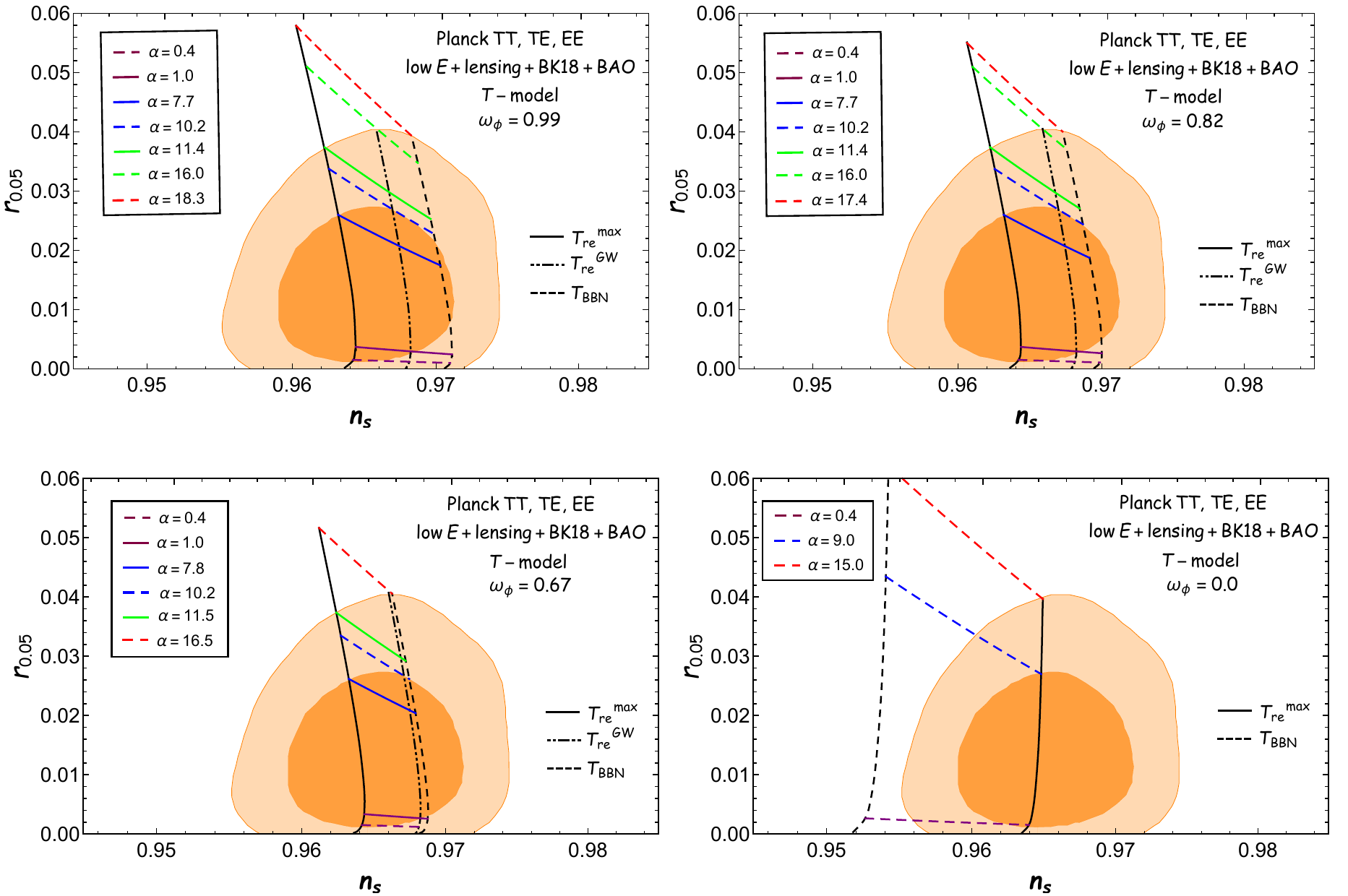}
    \caption{\em Prediction of $\alpha$-attractor T-model for different $(\alpha, w_{\phi})$. Detailed descriptions are the same as illustrated in Fig.\ref{nsrE}}
          \label{nsrT}
          \end{center}
      \end{figure*}
        \begin{figure*}
         \begin{center}
\includegraphics[width=0014.50cm]
          {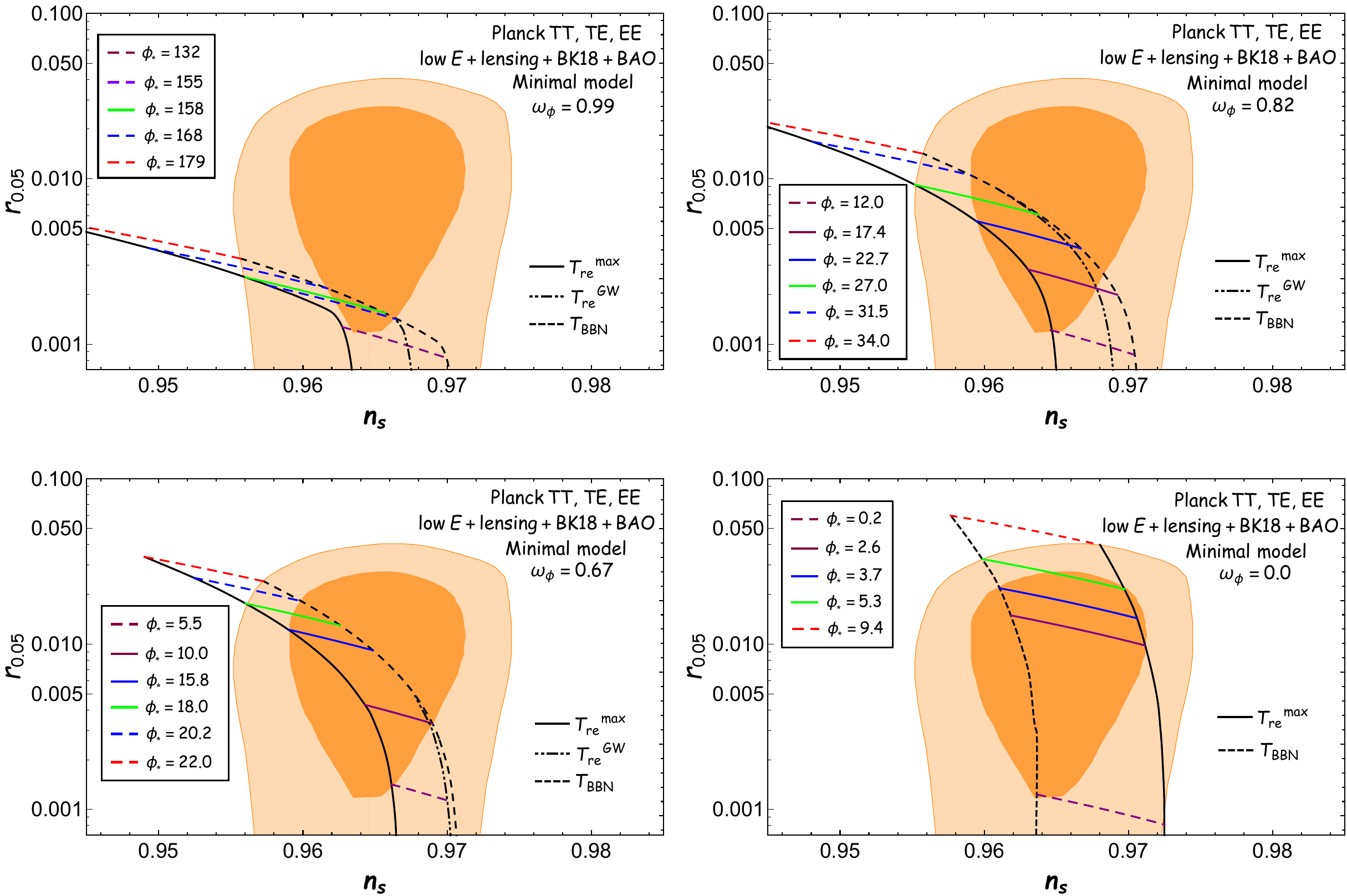}
          \caption{\em Prediction of minimal model for different values of $(\phi_*, w_{\phi})$. Detailed descriptions are the same as illustrated in Fig.\ref{nsrE}}
          \label{nsrminimal}
          \end{center}
      \end{figure*}
           \begin{figure*}
         \begin{center}
\includegraphics[width=0016.50cm]
          {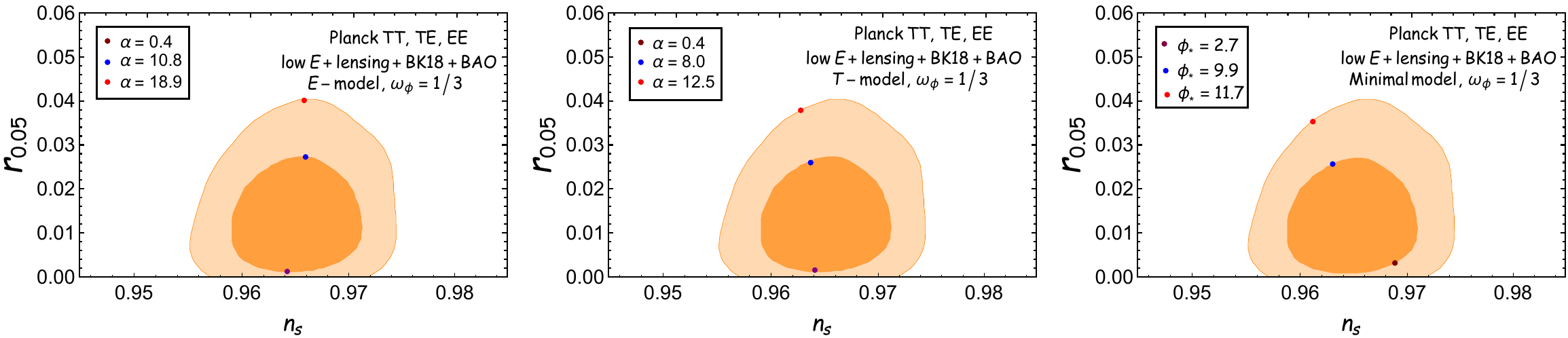}
          \caption{\em Illustrations of the restriction on the different inflationary models from recent BICEP/$Keck$ data for a specific value of $w_\phi=1/3$. }
          \label{w13}
          \end{center}
      \end{figure*}
     \begin{table}[t!]
      \scriptsize{
	\caption{Limiting values of ($N_{\rm k}, H_{\rm end}, \alpha$): attractor model}\label{infresE}
 E-model\\[.15cm]
	 \begin{tabular}{|p{.6cm}|p{.7cm}|p{1.4cm}|p{1.1cm}|p{.7cm}|p{1.2cm}|p{1.2cm}|}
\hline
\,$w_\phi$&\multicolumn{3}{c|}{1 $\sigma +$ PGWs}&\multicolumn{3}{c|}{2 $\sigma +$ PGWs}\\
\cline{2-7}
~&$\alpha_{\rm max}$&$H_{\rm end}^{\rm min}$ (GeV)&$N_{\rm k}$&$\alpha_{\rm max}$&$H_{\rm end}^{\rm min}$ (GeV)&$\quad\quad N_{\rm k}$\\
\hline
0.99&10.2&$2.6\times10^{12}$&64.7,55.6&16.0&$1.6\times10^{12}$&65.4,55.1\\
0.82&10.4&$3.5\times10^{12}$&64.5,55.6&17.5&$2.1\times10^{12}$&65.1,55.1\\
0.67&10.6&$4.4\times10^{12}$&64.2,55.7&19.5&$3.1\times10^{12}$&64.8,55.1\\
\hline
\end{tabular}\\[.2cm]
T-model\\[.15cm]
 \begin{tabular}{|p{.6cm}|p{.7cm}|p{1.4cm}|p{1.1cm}|p{.7cm}|p{1.2cm}|p{1.2cm}|}
\hline
\,$w_\phi$&\multicolumn{3}{c|}{1 $\sigma +$ PGWs}&\multicolumn{3}{c|}{2 $\sigma +$ PGWs}\\
\cline{2-7}
~&$\alpha_{\rm max}$&$H_{\rm end}^{\rm min}$ (GeV)&$N_{\rm k}$&$\alpha_{\rm max}$&$H_{\rm end}^{\rm min}$ (GeV)&$N_{\rm k}$\\
\hline
0.99&10.2&$2.9\times10^{12}$&64.7,55.7&16.0&$1.9\times10^{12}$&65.3,55.1\\
0.82&10.2&$3.0\times10^{12}$&64.6,55.6&16.0&$2.0\times10^{12}$&65.2,55.1\\
0.67&10.2&$3.3\times10^{12}$&64.5,55.6&16.5&$2.3\times10^{12}$&65.1,55.1\\
\hline
\end{tabular}}
\end{table}

\begin{table}[t!]
\scriptsize{
	\caption{Limiting values of ($N_{\rm k}, H_{\rm end}, \alpha, \phi_*$) }\label{infresE0}
For $w_\phi=0.0$\\[.1cm]
	 \begin{tabular}{|p{1.0cm}|p{.7cm}|p{1.3cm}|p{1.1cm}|p{.70cm}|p{1.3cm}|p{1.05cm}|}
\hline
Model&\multicolumn{3}{c|}{1 $\sigma $}&\multicolumn{3}{c|}{2 $\sigma $}\\
\cline{2-7}
&$\alpha, \phi_*$ (max)&$H_{\rm end}^{\rm min}$ (GeV)&$N_{\rm k}$&$\alpha,\phi_*$ (max)&$H_{\rm end}^{\rm min}$ (GeV)&$N_{\rm k}$\\
\hline
\quad E&13.0&$1.1\times10^{13}$&56.2,41.8&27.5&$1.1\times10^{13}$&56.4,41.5\\
\quad T&9.0&$8.6\times10^{12}$&56.4,41.5&15&$8.7\times10^{12}$&56.6,41.5\\
Minimal&5.3&$1.0\times10^{13}$&56.1,41.5&$9.4$&$9.5\times10^{12}$&56.5,41.5\\
\hline
\end{tabular}\\[.2cm]
For $w_\phi=1/3$\\[.15cm]
	 \begin{tabular}{|p{1.0cm}|p{.7cm}|p{1.3cm}|p{1.1cm}|p{.70cm}|p{1.3cm}|p{1.05cm}|}
\hline
Model&\multicolumn{3}{c|}{1 $\sigma $}&\multicolumn{3}{c|}{2 $\sigma $}\\
\cline{2-7}
&$\alpha, \phi_*$ (max)&$H_{\rm end}^{\rm min}$ (GeV)&$N_{\rm k}$&$\alpha,\phi_*$ (max)&$H_{\rm end}^{\rm min}$ (GeV)&$N_{\rm k}$\\
\hline
\quad E&10.8&$7.6\times10^{12}$&56.7,55.3&18.9&$6.7\times10^{12}$&56.9,54.9\\
\quad T&9.0&$5.7\times10^{12}$&56.8,55.4&12.5&$5.1\times10^{12}$&57.0,54.9\\
Minimal&9.9&$4.0\times10^{12}$&57,55.6&$11.7$&$4\times10^{12}$&57.0,54.2\\
\hline
\end{tabular}}
\end{table}

\begin{table*}
\begin{center}
\scriptsize{
\caption{Liminting values of ($N_{\rm k}, H_{\rm end}, \phi_*$) for minimal model}\label{minimalinf}
\begin{tabular}{|p{.6cm}|p{1.4cm}|p{1.2cm}|p{1.5cm}|p{1.4cm}|p{1.8cm}|p{1.7cm}|p{1.4cm}|p{1.7cm}|p{1.5cm}|}
\hline
\,$w_\phi$&\multicolumn{3}{c|}{1 $\sigma +T_{\rm re}^{\rm max}$}&\multicolumn{3}{c|}{1 $\sigma +$ BBN+PGWs}&\multicolumn{3}{c|}{2 $\sigma +$ BBN+PGWs}\\
\cline{2-10}
&$\phi_{\star}^{\rm min}$ &$H_{\rm end}^{\rm max}$ (GeV)&\quad$N_k$&$\phi_{\star}^{\rm max}$ &$H_{\rm end}^{\rm min}$ (GeV)&\quad$N_{\rm k}^{\rm max}$&$\phi_{\star}^{\rm max}$ &$H_{\rm end}^{\rm min}$ (GeV)&\quad$N_{\rm k}$\\
\hline
0.99&$132.0$&$2.4\times10^{12}$&$74.5,57.4$&$168.0$&$4.0\times10^{5}$&$74.5,57.4$&$179.0$&$8.6\times 10^2$&$76.7,51.0$\\
0.82&$12.0$&$1.4\times10^{12}$&$67.5,56.8$&$31.5$&$5.8\times10^{10}$&$67.5,56.8$&$34.0$&$5.2\times 10^{10}$&$67.7,52.6$\\
0.67&$5.5$&$5.1\times10^{12}$&$66.2,55.6$&$20.2$&$1.9\times10^{11}$&$66.2,55.6$&$22.0$&$1.4\times 10^{11}$&$66.4,53.3$\\
\hline
\end{tabular}}
\end{center}
\end{table*}
So far, we have analyzed different sources of constraints from observation and theory and their possible direct and indirect impacts on inflation. We considered three plateau types of inflation models, namely $\alpha$-attractor E, T, and minimal models, and two possible reheating scenarios depending on the inflaton-radiation coupling. 
Before we embark on discussing the results, let us summarize the main inputs we are considering:
\begin{itemize}
    \item  The combined observation data of the latest PLANCK 2018 and BICEP/$Keck$  put stringent constraints on inflationary large-scale observable, namely, scalar spectral index $n_s$ and tensor to scalar ratio $r$. Such constraints directly  impact the possible range of inflationary parameters, such as effective mass, the potential height ($\Lambda$), and the potential's nature at the end of the inflation.   
    \item Subsequent reheating causes inflaton to decay into radiation through different decay channels are under consideration. Non-trivial reheating dynamics supplemented with the post-reheating entropy conservation relates different observable and the reheating parameters though Eq.\ref{entropy-conservation}, and that  immediately gives us additional constraints on the inflaton model through the maximum ($T_{\rm re} \sim 10^{15}$ GeV) and minimum ($T_{\rm re} \sim 4$ MeV) reheating temperature in the $(n_s,r)$ plane. The bound on reheating temperature will lead to bound on the inflaton-radiation couplings $(h, \tilde{g} = g/m_{\phi}^{\rm end})$.  
    \item PLANCK, BICEP/$Keck$ usually measures large-scale inflationary fluctuations. Interestingly the small-scale inflationary tensor fluctuations (PGWs) have been observed to play an interesting role in further constraining the possible range of reheating temperature. If the reheating period is prolonged, and reheating equation of state $w_{\phi} > 1/3$, the high-frequency gravitational waves acquire blue tilted spectrum $\Omega_{\rm GW}^{\rm k} \sim k^{n_{w_{\phi}}}$, where $n_{w_{\phi}}$ is the index of the spectrum and that may lead to larger lower bound on reheating temperature $T_{\rm re}^{\rm GW} > T_{\rm BBN}$. Such lower bounds naturally lead to tighter bounds on the inflaton-radiation couplings.
    \item In the perturbative framework, instantaneous reheating gives a natural upper bound on the inflaton-radiation coupling. However, generically such coupling modifies the inflaton potential at the loop level during inflation (CW constraints (CWc)). The loop-corrected inflaton potential should not disturb the inflation (see Fig.\ref{couplingcw}), which may modify the maximum reheating temperature and, consequently, the inflaton-radiation coupling parameter. We consider those bound throughout our analysis. 
    \item We further identify the nonperturbative constraints (NPc) on the inflaton-radiation coupling by employing broad parametric resonance condition taking into account the dynamical nature of resonance parameter $q(t)$ (see Eq.\ref{qt}). In the final results, we will take into account those as well. 
\end{itemize} 
 \begin{figure*}[t]
         \begin{center}
      \includegraphics[width=0017.50cm]
          {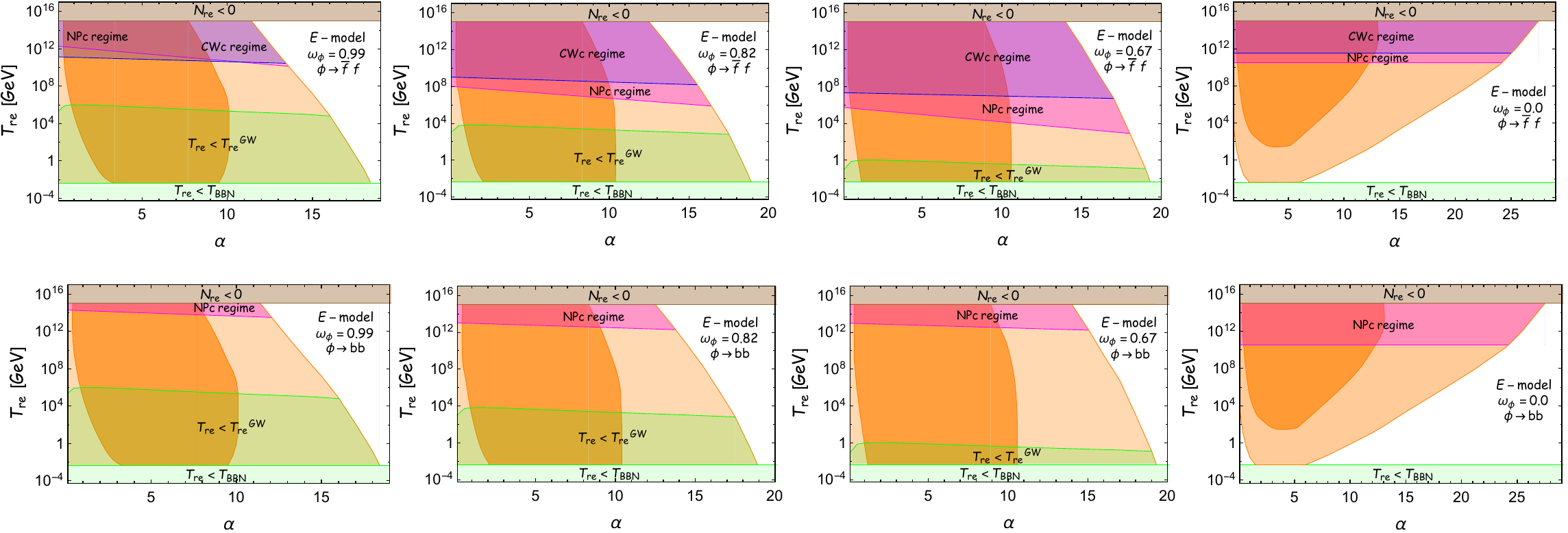}     \includegraphics[width=0017.50cm,height=07.0cm]
          {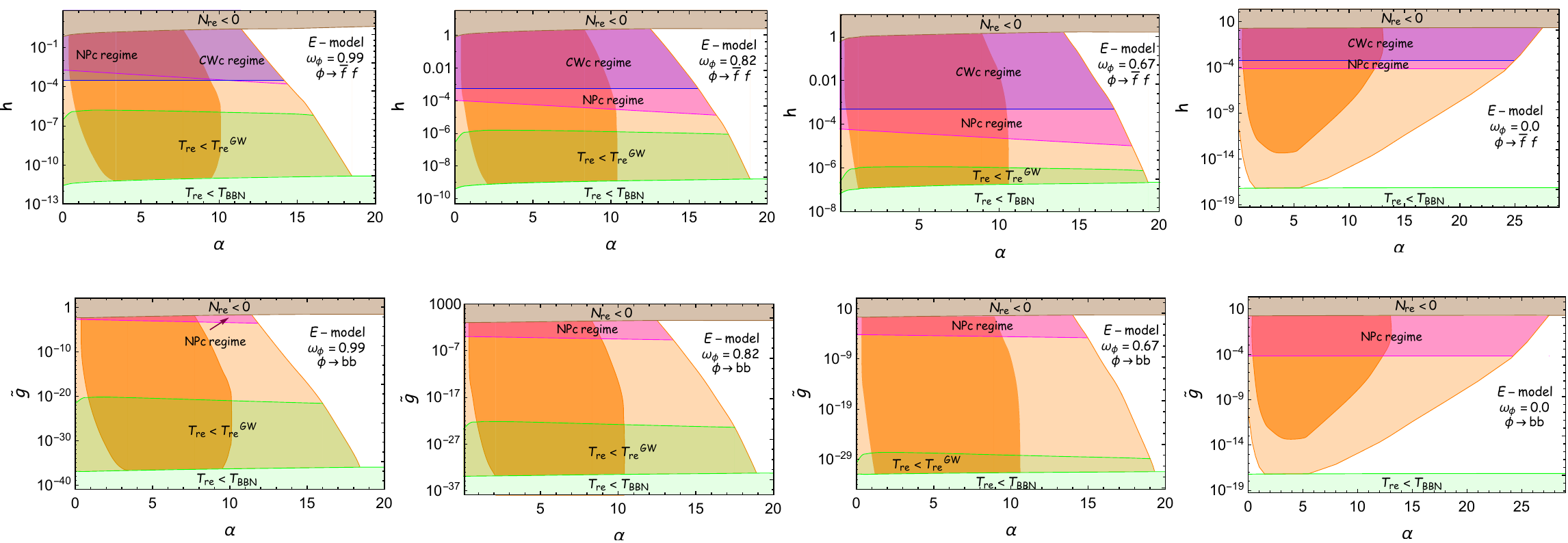}
          \caption{\em The impact of {\rm PLANCK 2018 + BICEP/Keck + PGW + BBN + CWc + NPc} on $T_{\rm re}$ and dimensionless inflaton-radiation coupling  ($h,\tilde{g}$) with respect to $\alpha$ for $\alpha$-attractor E-model. We take $w_{\phi} = (0,0.67,0.82,0.99)$. Deep orange shaded regions correspond to $1\sigma$ and light orange shaded regions correspond to $2\sigma$ bound imported from ($n_s-r$) plane. The deep and light green region indicates $T_{\rm re} < T_{\rm BBN}$ and $T_{\rm re} < T_{\rm re}^{\rm GW}$, respectively. Blue and magenta-shaded regions depict CWc and NPc. Gray-shaded region implies no reheating.}
          \label{Tree}
          \end{center}
      \end{figure*}
      \begin{table*}[t!]
      \scriptsize{
	\caption{Bounds on $(T_{re}, \Tilde{g},\,h)$ from PGWs+BBN+CWc+NPc with recent BICEP/$Keck$ data for $\alpha$-attractor E-model}\label{rehconE}}
I) Fermionic reheating ($\phi\to \bar{f}f$)\\[.1cm]
\begin{tabular}{|p{.55cm}|p{1.25cm}|p{2.3cm}|p{1.4cm}|p{2.2cm}|p{1.85cm}|p{2.35cm}|p{1.85cm}|p{2.35cm}|}
\hline
\,$w_\phi$&\multicolumn{2}{c|}{\scriptsize{1$\sigma$+BBN+PGWs+CWc}}&\multicolumn{2}{c|}{\scriptsize{1$\sigma$+BBN+PGWs+CWc+NPc}}&\multicolumn{2}{c|}{\scriptsize{2$\sigma$+BBN+PGWs+CWc}}&\multicolumn{2}{c|}{\scriptsize{2 $\sigma$+BBN+PGWs+CWc+NPc}}\\
\cline{2-9}
~&${T}_{\rm re}$\scriptsize{(GeV)}&\quad$h$&$ T_{\rm re}$ \scriptsize{(GeV)}&\quad$h$&$T_{\rm re}$ \scriptsize{(GeV)}&$\quad h$&$T_{\rm re}$ \scriptsize{(GeV)}&$\quad h$\\
\hline
0.99&$10^{11},10^5$&$3\times10^{-4},10^{-6}$&$10^{12},10^5$&$10^{-3},10^{-6}$&$10^{11},7\times 10^4$&$3\times10^{-4},7\times10^{-7}$&$10^{12},7\times 10^4$&$10^{-3},8\times 10^{-7}$\\
\hline
0.82&$10^{9},10^3$&$3\times10^{-4},10^{-6}$&$10^{8},10^3$&$7\times10^{-5},10^{-6}$&$10^{9},7\times 10^2$&$3\times 10^{-4},10^{-6}$&$10^{8},7\times 10^2$&$7\times10^{-5}, 10^{-6}$\\
\hline
0.67&$10^{7},0.3$&$3\times10^{-4},5\times10^{-7}$&$5\times10^{5},0.3$&$6\times10^{-5},5\times10^{-7}$&$10^{7},0.1$&$5\times10^{-4},4\times 10^{-7}$&$5\times10^{5},0.1$&$6\times10^{-5}, 4\times10^{-7}$\\
\hline
1/3&$10^{6},0.004$&$3\times 10^{-4},10^{-9}$&$10^{5},0.004$&$6\times10^{-5},10^{-9}$&$10^{6},0.004$&$3\times10^{-4},10^{-9}$&$10^{5},0.004$&$6\times10^{-5}, 10^{-9}$\\
\hline
0.0&$10^{11},10$&$5\times10^{-4},10^{-14}$&$10^{10},10$&$6\times10^{-5},10^{-14}$&$10^{11},0.004$&$5\times10^{-4}, 10^{-17}$&$10^{10},0.004$&$6\times10^{-5},10^{-17}$\\
\hline
\end{tabular}\\[.2cm]
II) Bosonic reheating ($\phi\to bb$)\\[.1cm]
	\begin{tabular}{|p{.6cm}|p{1.5cm}|p{1.6cm}|p{1.7cm}|p{2.6cm}|p{1.85cm}|p{1.9cm}|p{1.85cm}|p{2.4cm}|}
\hline
\,$w_\phi$&\multicolumn{2}{c|}{\scriptsize{1$\sigma$+BBN+PGWs+CWc}}&\multicolumn{2}{c|}{\scriptsize{1$\sigma$+BBN+PGWs+CWc+NPc}}&\multicolumn{2}{c|}{\scriptsize{2$\sigma$+BBN+PGWs+CWc}}&\multicolumn{2}{c|}{\scriptsize{2$\sigma$+BBN+PGWs+CWc+NPc}}\\
\cline{2-9}
~&${T}_{\rm re}$\scriptsize{(GeV)}&\quad$\tilde g$&$ T_{\rm re}$ \scriptsize{(GeV)}&\quad$\tilde g$&$T_{\rm re}$ (GeV)&$\quad \tilde g$&$T_{\rm re}$ \scriptsize{(GeV)}&$\quad\tilde g$\\
\hline
0.99&$10^{15},10^5$&$10^{-2},10^{-21}$&$10^{14},10^5$&$10^{-3},10^{-21}$&$10^{15},7\times 10^4$&$10^{-2},10^{-22}$&$10^{14},7\times 10^4$&$10^{-3},10^{-22}$\\
\hline
0.82&$10^{15},10^3$&$0.3,10^{-23}$&$10^{13},10^3$&$7\times10^{-5},10^{-23}$&$10^{15},7\times 10^2$&$0.3,3\times 10^{-24}$&$10^{13},7\times 10^2$&$7\times10^{-5},3\times 10^{-24}$\\
\hline
0.67&$10^{15},0.3$&$0.4,10^{-29}$&$4\times10^{12},0.3$&$6\times10^{-5},10^{-29}$&$10^{15},0.1$&$0.5,3\times 10^{-29}$&$4\times10^{12},0.1$&$6\times10^{-5},3\times10^{-29}$\\
\hline
1/3&$10^{15},0.004$&$0.5,10^{-27}$&$10^{11},0.004$&$6\times10^{-5},10^{-27}$&$10^{15},0.004$&$0.5,10^{-27}$&$10^{11},0.004$&$6\times10^{-5}, 10^{-27}$\\
\hline
0.0&$10^{15},10$&$0.5,10^{-14}$&$10^{10},10$&$6\times10^{-5},10^{-14}$&$10^{15},0.004$&$0.5,10^{-17}$&$10^{10},0.004$&$6\times10^{-5}, 10^{-17}$\\
\hline
\end{tabular}
\end{table*}
        \begin{figure*}[t!]
         \begin{center}
      \includegraphics[width=0017.50cm]
          {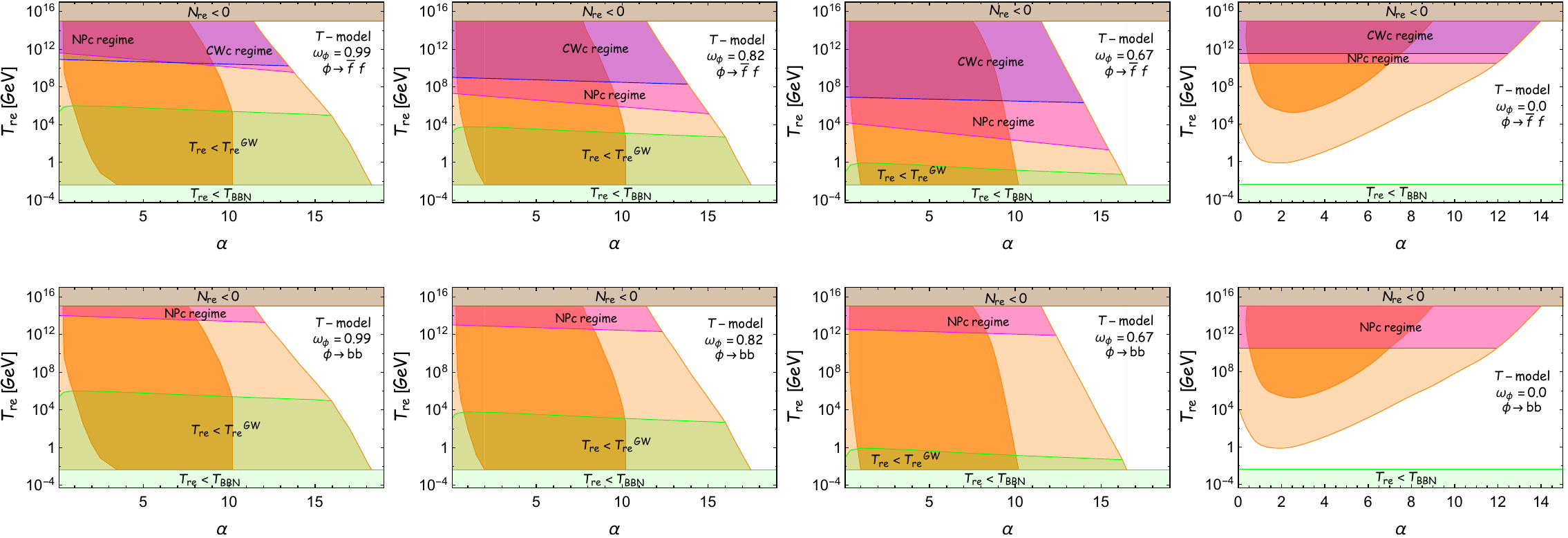}       \includegraphics[width=0017.50cm,height=07.cm]
          {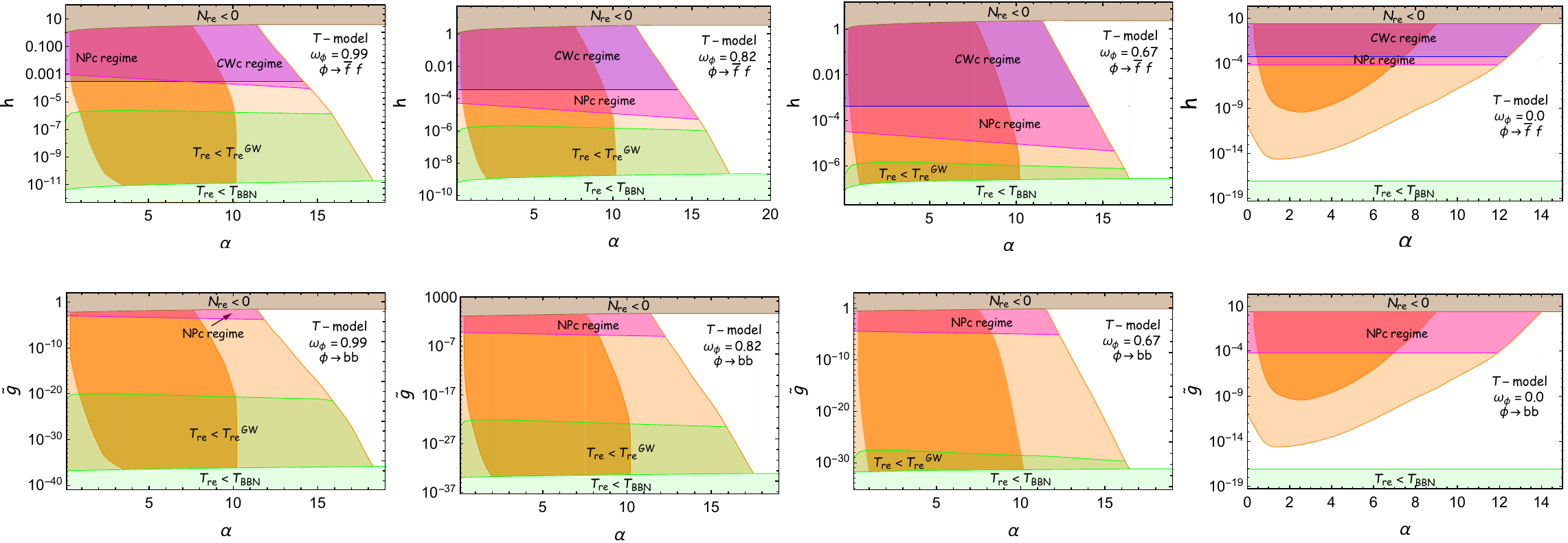}
          \caption{\em The impact of {\rm PLANCK 2018 + BICEP/Keck + PGW + BBN + CWc + NPc} on $T_{\rm re}$ and dimensionless inflaton-radiation coupling  ($h,\tilde{g}$) with respect to $\alpha$ for $\alpha$-attractor T-model. We take $w_{\phi} = (0,0.67,0.82,0.99)$. Deep orange shaded regions correspond to $1\sigma$ and light orange shaded regions correspond to $2\sigma$ bound imported from ($n_s-r$) plane. The deep and light green regions indicate $T_{\rm re} < T_{\rm BBN}$ and $T_{\rm re} < T_{\rm re}^{\rm GW}$, respectively. Blue and magenta-shaded regions depict CWc and NPc. Gray-shaded region implies no reheating.}
          \label{Tret}
          \end{center}
      \end{figure*}
\begin{table*}[t!]
\scriptsize{
\caption{Bounds on ($T_{re}, \Tilde{g},\,h$) for $\alpha$-attractor T-model }\label{rehconT}}
I) Fermionic reheating ($\phi\to \bar{f}f$)\\[.1cm]
\begin{tabular}{|p{.55cm}|p{1.25cm}|p{2.3cm}|p{1.4cm}|p{2.2cm}|p{1.85cm}|p{2.35cm}|p{1.85cm}|p{2.35cm}|}
\hline
\,$w_\phi$&\multicolumn{2}{c|}{1 $\sigma$+BBN+PGWs+CWc}&\multicolumn{2}{c|}{1 $\sigma$+BBN+PGWs+CWc+NPc}&\multicolumn{2}{c|}{2 $\sigma$+BBN+PGWs+CWc}&\multicolumn{2}{c|}{2 $\sigma$+BBN+PGWs+CWc+NPc}\\
\cline{2-9}
~&${T}_{\rm re}$ (GeV)&\quad$h$&$ T_{\rm re}$ (GeV)&\quad$h$&$T_{\rm re}$ (GeV)&$\quad h$&$T_{\rm re}$ (GeV)&$\quad h$\\
\hline
0.99&$10^{11},10^5$&$3\times10^{-4},10^{-6}$&$10^{12},10^5$&$10^{-3},10^{-6}$&$10^{11},7\times 10^4$&$3\times10^{-4},7\times10^{-7}$&$10^{12},7\times 10^4$&$10^{-3},8\times 10^{-7}$\\
\hline
0.82&$10^{9},10^3$&$3\times10^{-4},10^{-6}$&$10^{8},10^3$&$7\times10^{-4},10^{-6}$&$10^{9},7\times 10^2$&$3\times 10^{-4},10^{-6}$&$10^{8},7\times 10^2$&$7\times10^{-5}, 10^{-6}$\\
\hline
0.67&$10^{7},0.3$&$3\times10^{-4},5\times10^{-7}$&$5\times10^{5},0.3$&$6\times10^{-5},5\times10^{-7}$&$10^{7},0.1$&$5\times10^{-4},4\times 10^{-7}$&$5\times10^{5},0.1$&$6\times10^{-5}, 4\times10^{-7}$\\
\hline
1/3&$10^{6},0.004$&$3\times 10^{-4},10^{-9}$&$10^{5},0.004$&$6\times10^{-5},10^{-9}$&$10^{6},0.004$&$3\times10^{-4},10^{-9}$&$10^{5},0.004$&$6\times10^{-5}, 10^{-9}$\\
\hline
0.0&$10^{11},10^5$&$5\times10^{-4},10^{-10}$&$10^{10},10^5$&$6\times10^{-5},10^{-10}$&$10^{11},0.6$&$5\times10^{-4},10^{-15}$&$10^{10},0.6$&$6\times10^{-5},10^{-15}$\\
\hline
\end{tabular}\\[.2cm]
 II) Bosonic reheating ($\phi\to bb$)\\[.1cm]
	\begin{tabular}{|p{.6cm}|p{1.5cm}|p{1.6cm}|p{1.7cm}|p{2.6cm}|p{1.85cm}|p{1.9cm}|p{1.85cm}|p{2.4cm}|}
\hline
\,$w_\phi$&\multicolumn{2}{c|}{\scriptsize{1$\sigma$+BBN+PGWs+CWc}}&\multicolumn{2}{c|}{\scriptsize{1$\sigma$+BBN+PGWs+CWc+NPc}}&\multicolumn{2}{c|}{\scriptsize{2$\sigma$+BBN+PGWs+CWc}}&\multicolumn{2}{c|}{\scriptsize{2$\sigma$+BBN+PGWs+CWc+NPc}}\\
\cline{2-9}
~&${T}_{\rm re}$ \scriptsize{(GeV)}&\quad$\tilde g$&$ T_{\rm re}$ \scriptsize{(GeV)}&\quad$\tilde g$&$T_{\rm re}$ \scriptsize{(GeV)}&$\quad \tilde g$&$T_{\rm re}$\scriptsize{ (GeV)}&$\quad\tilde g$\\
\hline
0.99&$10^{15},10^5$&$10^{-2},10^{-21}$&$10^{14},10^5$&$10^{-3},10^{-21}$&$10^{15},10^5$&$0.02,10^{-21}$&$10^{14},10^5$&$10^{-3},10^{-22}$\\
\hline
0.82&$10^{15},10^3$&$0.3,10^{-23}$&$2\times10^{13},10^3$&$5\times10^{-5},10^{-23}$&$10^{15},4\times 10^2$&$0.3,3\times 10^{-24}$&$10^{13},4\times 10^2$&$7\times10^{-5},10^{-24}$\\
\hline
0.67&$10^{15},0.1$&$0.4,10^{-29}$&$4\times10^{12},0.1$&$6\times10^{-5},10^{-29}$&$10^{15},0.05$&$0.5,10^{-29}$&$4\times10^{12},0.05$&$6\times10^{-5},10^{-29}$\\
\hline
1/3&$10^{15},0.004$&$0.5,10^{-27}$&$10^{11},0.004$&$6\times10^{-5},10^{-27}$&$10^{15},0.004$&$0.5,10^{-27}$&$10^{11},0.004$&$6\times10^{-5}, 10^{-27}$\\
\hline
0.0&$10^{15},10^5$&$0.5,10^{-10}$&$10^{10},10^5$&$6\times10^{-5},10^{-10}$&$10^{15},0.6$&$0.5,2\times 10^{-15}$&$4\times 10^{10},0.6$&$6\times10^{-5},10^{-15}$\\
\hline
\end{tabular}
\end{table*}
\begin{figure*}
         \begin{center}
      \includegraphics[width=0017.50cm]
          {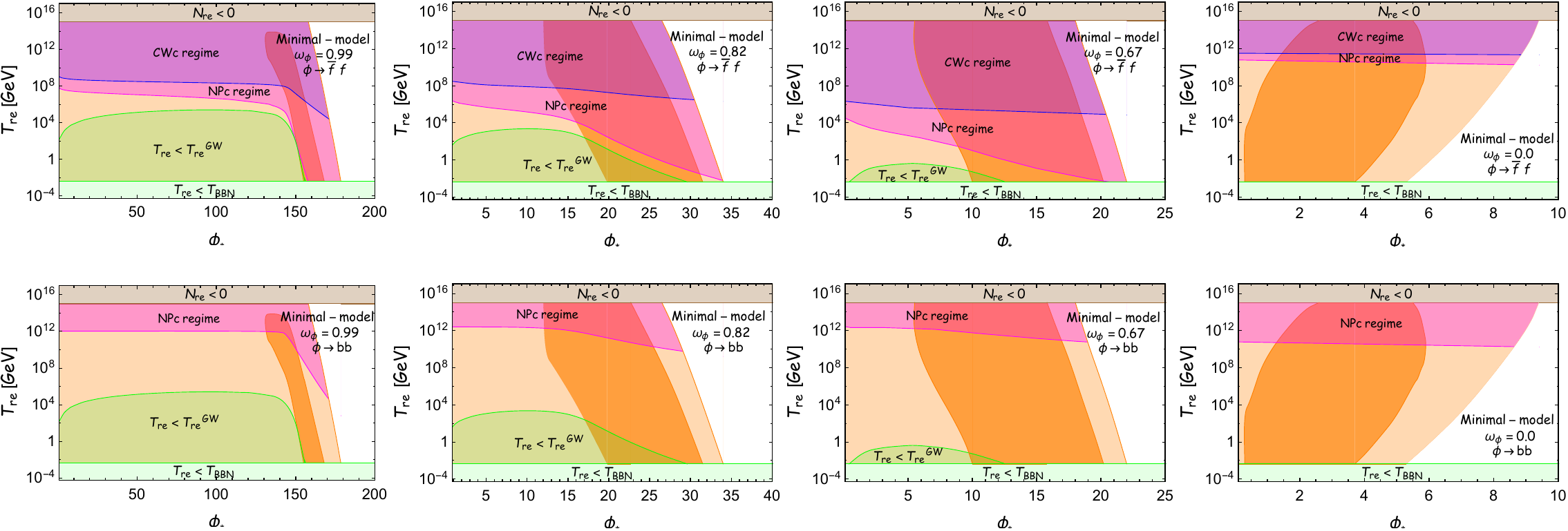}        \includegraphics[width=0017.50cm,height=07.cm]
          {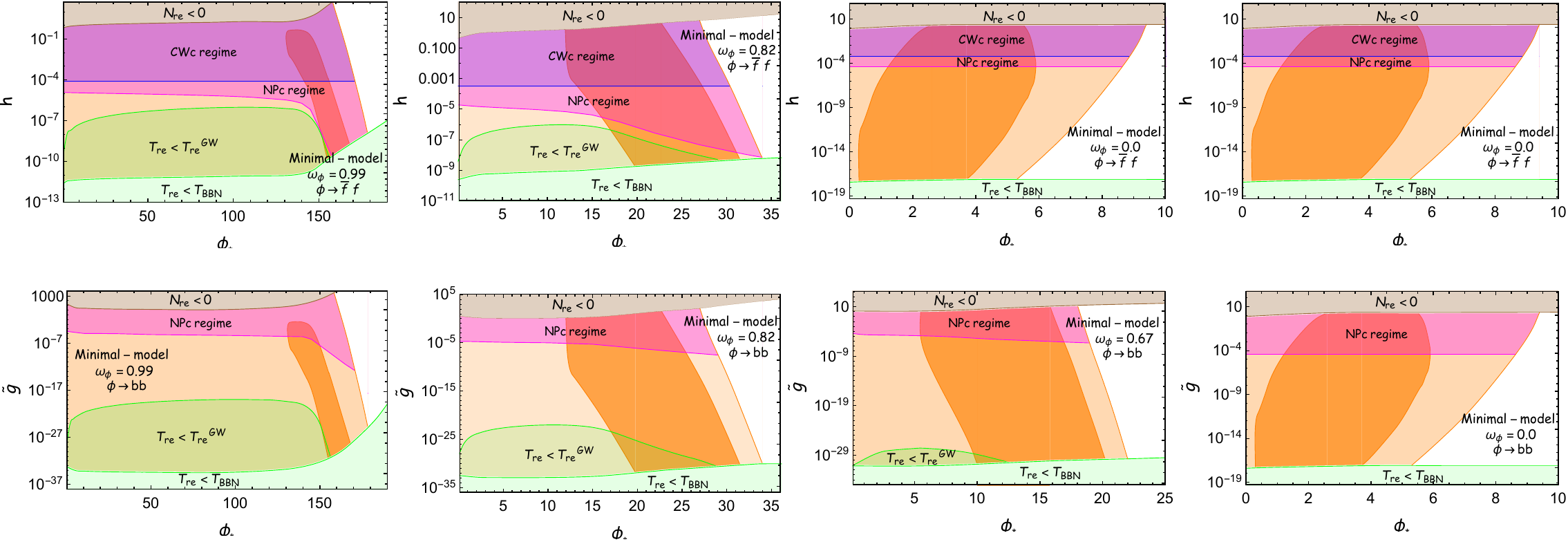}
          \caption{\em The impact of {\rm PLANCK 2018 + BICEP/Keck + PGW + BBN + CWc + NPc} on $T_{\rm re}$ and dimensionless inflaton-radiation coupling  ($h,\tilde{g}$) with respect to $\phi_\star$ for the minimal model. We take $w_{\phi} = (0,0.67,0.82,0.99)$. Deep orange shaded regions correspond to $1\sigma$, and the light orange shaded regions correspond to $2\sigma$ bound imported from ($n_s-r$) plane. The deep and light green regions indicate $T_{\rm re} < T_{\rm BBN}$ and $T_{\rm re} < T_{\rm re}^{\rm GW}$, respectively. Blue and magenta-shaded regions depict CWc and NPc. Gray-shaded region implies no reheating.}
          \label{Trem}
          \end{center}
      \end{figure*}
 \begin{table*}[t!]
\scriptsize{
\caption{Bounds on ($T_{re}, \Tilde{g},\,h$) for  Minimal model}\label{minimalres}
I) Fermionic reheating ($\phi\to \bar{f}f$)\\[.1cm]
\begin{tabular}{|p{.55cm}|p{1.3cm}|p{2.1cm}|p{1.4cm}|p{2.2cm}|p{1.85cm}|p{2.35cm}|p{1.85cm}|p{2.35cm}|}
\hline
\,$w_\phi$&\multicolumn{2}{c|}{1 $\sigma$+BBN+PGWs+CWc}&\multicolumn{2}{c|}{1 $\sigma$+BBN+PGWs+CWc+NPc}&\multicolumn{2}{c|}{2 $\sigma$+BBN+PGWs+CWc}&\multicolumn{2}{c|}{2 $\sigma$+BBN+PGWs+CWc+NPc}\\
\cline{2-9}
~&${T}_{\rm re}$ (GeV)&\quad$h$&$ T_{\rm re}$ (GeV)&\quad$h$&$T_{\rm re}$ (GeV)&$\quad h$&$T_{\rm re}$ (GeV)&$\quad h$\\
\hline
0.99&$10^{9},0.004$&$7\times10^{-4},10^{-10}$&$10^{8},0.004$&$3\times10^{-5},10^{-10}$&$10^{9},0.004$&$7\times10^{-5},10^{-10}$&$10^{8},0.004$&$3\times10^{-5},10^{-10}$\\
\hline
0.82&$10^{9},0.004$&$3\times10^{-4},10^{-9}$&$10^{6},0.004$&$2\times10^{-5},10^{-9}$&$10^{9},0.004$&$3\times10^{-4},10^{-9}$&$10^{6},0.004$&$2\times10^{-5},10^{-9}$\\
\hline
0.67&$10^{5},0.004$&$3\times10^{-4},10^{-9}$&$20,0.004$&$2\times10^{-6},10^{-9}$&$10^{6},0.004$&$3\times10^{-4},10^{-9}$&$10^{4},0.004$&$2\times10^{-5},10^{-9}$\\
\hline
1/3&$10^{6},0.004$&$3\times 10^{-4},10^{-9}$&$10^{5},0.004$&$6\times10^{-5},10^{-9}$&$10^{6},0.004$&$3\times10^{-4},10^{-9}$&$10^{5},0.004$&$6\times10^{-5}, 10^{-9}$\\
\hline
0.0&$10^{11},0.004$&$4\times10^{-4},10^{-17}$&$10^{10},0.004$&$4\times10^{-5},10^{-17}$&$10^{11},0.004$&$4\times10^{-4},10^{-17}$&$10^{10},0.004$&$4\times10^{-5},10^{-17}$\\
\hline
\end{tabular}}\\[.2cm]
 II) Bosonic reheating ($\phi\to bb$)\\[.1cm]
	\begin{tabular}{|p{.6cm}|p{1.5cm}|p{1.6cm}|p{1.7cm}|p{2.6cm}|p{1.85cm}|p{1.9cm}|p{1.85cm}|p{2.4cm}|}
\hline
\,$w_\phi$&\multicolumn{2}{c|}{1 $\sigma$+BBN+PGWs+CWc}&\multicolumn{2}{c|}{1 $\sigma$+BBN+PGWs+CWc+NPc}&\multicolumn{2}{c|}{2 $\sigma$+BBN+PGWs+CWc}&\multicolumn{2}{c|}{2 $\sigma$+BBN+PGWs+CWc+NPc}\\
\cline{2-9}
~&${T}_{\rm re}$ (GeV)&\quad$\tilde g$&$ T_{\rm re}$ (GeV)&\quad$\tilde g$&$T_{\rm re}$ (GeV)&$\quad \tilde g$&$T_{\rm re}$ (GeV)&$\quad\tilde g$\\
\hline
0.99&$10^{14},0.004$&$10^{-3},10^{-32}$&$10^{12},0.004$&$4\times10^{-5},10^{-32}$&$10^{15},0.004$&$2\times10^3,10^{-32}$&$10^{12},0.004$&$4\times10^{-5},10^{-32}$\\
\hline
0.82&$10^{15},0.004$&$30,10^{-31}$&$10^{12},0.004$&$2\times10^{-5},10^{-31}$&$10^{15},0.004$&$10^2, 10^{-31}$&$10^{12},0.004$&$2\times10^{-5},10^{-31}$\\
\hline
0.67&$10^{15},0.004$&$0.1,10^{-31}$&$10^{12},0.004$&$2\times10^{-5},10^{-31}$&$10^{15},0.004$&$0.1,10^{-31}$&$10^{12},0.004$&$2\times10^{-5},10^{-31}$\\
\hline
1/3&$10^{15},0.004$&$0.5,10^{-27}$&$10^{11},0.004$&$6\times10^{-5},10^{-27}$&$10^{15},0.004$&$0.5,10^{-27}$&$10^{11},0.004$&$6\times10^{-5}, 10^{-27}$\\
\hline
0.0&$10^{15},0.004$&$0.5,10^{-17}$&$10^{10},0.004$&$4\times10^{-5},10^{-17}$&$10^{15},0.004$&$0.5,10^{-17}$&$10^{10},0.004$&$4\times10^{-5},10^{-17}$\\
\hline
\end{tabular}
\end{table*}
\noindent
{\bf PLANCK + BICEP/$Keck$ + BBN + PGWs + CWc + NPc constraints on inflation}: \\ 
First, we display our results in $(n_{\rm s}-r)$ plane and compare it with the latest observational data (see Figs.\ref{nsrE}-\ref{w13}) for the different inflationary models with five reheating equation of states $w_\phi=(0,\,1/3,\,0.67,\,0.82,\,0.99)$. In those plots, the color code for $1\sigma$(deep orange) and $2\sigma$ (light orange) regions are given by the combined data of Planck and BICEP/$Keck$. We use the same color code in subsequent plots to estimate the model parameters' $1\sigma$ and $2\sigma$ bounds. To make our representation clear, in Figs.\ref{nsrE}-\ref{w13}, we did not include CW-perturbative and nonperturbative constraints but incorporated them in all the subsequent plots. Lower limits on the $T_{\rm re}$ are set by either $T_{\rm BBN}$ or $T^{\rm GW}_{\rm re}$ depending on the model and $w_{\phi}$. 
The $T_{\rm re}^{\rm max}$ value differs depending on the decay channel for both  perturbative and nonperturbative considerations. 
For example, if we ignore both CW-perturbative bound and nonperturbative effect, for $\phi\to \bar{f}{f}$ reheating  instantaneous reheating sets the maximum reheating temperature $T_{\rm  re}^{\rm max} \sim 10^{15}$ GeV. Otherwise, it gets modified. However, ignoring the nonperturbative effect, CW-perturbative bound does not give additional constraint for bosonic reheating, and for that instantaneous reheating ($N_{re}\to0$) sets the value of $T_{\rm re}^{\rm max} \sim 10^{15}$ GeV.
It can be further observed that for stiff EoS $w_\phi>1/3$, $T_{\rm re}^{\rm min}$ set the maximum bound on the inflaton potential parameters $\alpha, \phi_*$, and for $w_\phi< 1/3$, it is $T_{\rm re}^{\rm max}$ which sets the bound.

As an example, $w_\phi=0$, the upper limit of the potential parameters $\alpha, \phi_*$ are  $(13.0,\,9.0,\,5.3)$ ($1\sigma$ bound) and $(27.5,\,15.0,\,9.4)$ ($2\sigma$ bound) for attractor E, T, and minimal model respectively (see Table-\ref{infresE0}). For higher EoS $w_\phi=(0.67,\,0.82,\,0.99)$, the restriction of $\alpha(\phi_*)$ are given in Tables-\ref{infresE} and \ref{minimalinf}. \\
 Once the maximum values of $(\alpha, \phi_*)$ are fixed, the associated prediction of inflationary energy scale and the e-folding number $N_{\rm k}$ can be computed. 
For the attractor models with stiff inflaton equation of state $w_\phi>1/3$, the predicted value of $(H_{\rm end}^{\rm min}\sim 10^{12}~{\rm GeV}, N_k^{\rm max} \sim 65)$ (both E and T-model) turned out to be nearly independent of $w_{\rm \phi}$ and the minimum possible e-folding number correspond to instantaneous reheating $N^{\rm min}_k\sim 55$ which is also independent of $w_\phi$. On the other hand, for the minimal plateau model, however, $H_{\rm end}^{\rm min}$ varies within the range $(10^{11}-10^5)$ GeV ($1\sigma$ bound) and $(10^{11}-10^3)\}$ GeV ($2\sigma$ bound) for $w_\phi$ varying from $0.67\to 0.99$. 
For the same range of $w_{\phi}$, $N_{\rm k}^{\rm max}$ value varies from $66\to 77$. On the other hand if, $w_\phi=0$, $ H_{\rm end}^{\rm min} \sim 10^{13}$ GeV, and $N_{\rm k}^{\rm max} \sim 56$ irrespective of the models under consideration. We summarize all the bounds of inflationary parameters in Tables-\ref{infresE}, \ref{infresE0}, and \ref{minimalinf}.\\
In the above, we discussed mainly the constraints on inflation parameters and associated prediction. However, for a complete understanding of the nature of inflation, we now consider the impact of the above constraints on the reheating history, which can directly constrain the inflaton-radiation coupling parameters $(h,\tilde g)$. Along with those, we include all the theoretical bounds obtained from both CWc and NPc discussed before. Let us briefly describe the color codes we used in our subsequent plots: 1) Deep green region is for $T_{\rm re}<T_{\rm BBN}$, 2) Light green region is associated with BBN constraints from PGWs, $T_{\rm re}<T_{\rm re}^{\rm GW}$, 3) Gray region depicts $N_{\rm re} < 0$, 4) Blue region termed as the CWc regime which indicates the restriction from CWc, 5) Magenta regime is associated with NPc, 6) Deep orange region indicates allowed coupling parameter space satisfying $1\sigma$ bound of latest $(n_s-r)$ data, and 7) Light orange region indicates allowed coupling parameter space satisfying $2\sigma$ bound of latest $(n_s-r)$ data.
\subsection{$\alpha$-attractor ${\rm E}~ \& ~{\rm T} $ models and constraints}
Constraints on the entire parameter space for attractor models can be summarized from Figs.\ref{nsrE},\ref{nsrT},\ref{w13},\ref{Tree},\ref{Tret}, and Tables \ref{infresE}, \ref{infresE0},\ref{rehconE},\ref{rehconT}. As stated earlier, we have discussed reheating scenarios described by two types of decay channels $\phi \to \bar{f}f$ and $\phi\to bb$. 
As argued before, the upper limit of the inflaton-radiation coupling should always be constrained by the fact that its quantum effect will not disturb the inflationary dynamics. If the bosonic channel dominates reheating, one loop-effective potential sets the upper bound on coupling $g \sim \mathcal{O}(10^{-1})$ and that fixes the $T_{\rm re}^{\rm max}$ value close to that of instantaneous reheating $ 10^{15}$ GeV. On the other hand, for the fermionic channel, such perturbative correction sets the upper bound on the coupling $h_{\rm max} \sim \mathcal{O}(10^{-4})$ and that leads to an upper limit on the reheating temperature, which depends on $w_\phi$. For instance, if $w_\phi=0$, $T_{\rm re}^{\rm max}\sim 10^{11}$ GeV for both the attractor models (see, for instance, the fourth plot of Figs.\ref{Tree},\ref{Tret}). 
For stiff inflaton EoS $w_\phi>1/3$, $T_{\rm re}^{\rm max}$  turns out to be nearly independent of $\alpha$ with the following values $T_{\rm re}^{\rm max} =(10^6,\,10^7,\,10^9,\,10^{11})$ GeV for $w_\phi=(0.33,\,0.67,\,0.82,\,0.99)$ respectively. \\
Unlike the upper limit, the lower limit on the inflaton-radiation coupling arises indirectly from the lower bound of the reheating temperature. 
If $w_\phi>0.60$, $T_{\rm re}^{\rm GW}$ always set the lowest reheating temperature, and the
value of $T_{\rm re}^{\rm GW}\sim (10^{-1},\,10^{3},\,10^{5})$ GeV for EoS $w_\phi=(0.67,\,0.82,\,0.99)$ respectively. 
Corresponding limiting values of couplings are tabulated in Tables-\ref{rehconE},\ref{rehconT}. When there is no restriction from PGWs ($w_\phi<1/3$), we expect $T_{\rm re}^{\rm min}\sim T_{\rm BBN}$ but that may not be true once we incorporate the PLACNK+BICEP/$Keck$ constraints. For two sample value, $w_\phi=0$ and $w_\phi=1/3$ , all numerical estimation of $T_{\rm re}^{\rm min}$ as well as $\tilde{g}_{min}/h_{min}$ shown in Tables-\ref{rehconE},\ref{rehconT}.\\
 As stated earlier, $T_{\rm re}^{\rm max}$ generally can be obtained  by CWc (Eq.\ref{critgh}); however, that may restrict from NPc (Eq.\ref{nonper2}) depending upon the model under consideration. As an example, for bosonic reheating ($\phi\to bb$ process) considering NPc the dimensionless coupling must be  $10^{-4} < \tilde{g} < 10^{-3}$, with the allowed range of $w_\phi=(0,\,1)$, and for $\phi\to \bar{f}f$  we have more or less same range, $(10^{-4}  < h < 10^{-3})$.
 \begin{figure*}[t]
         \begin{center}
\includegraphics[width=0017.50cm]
          {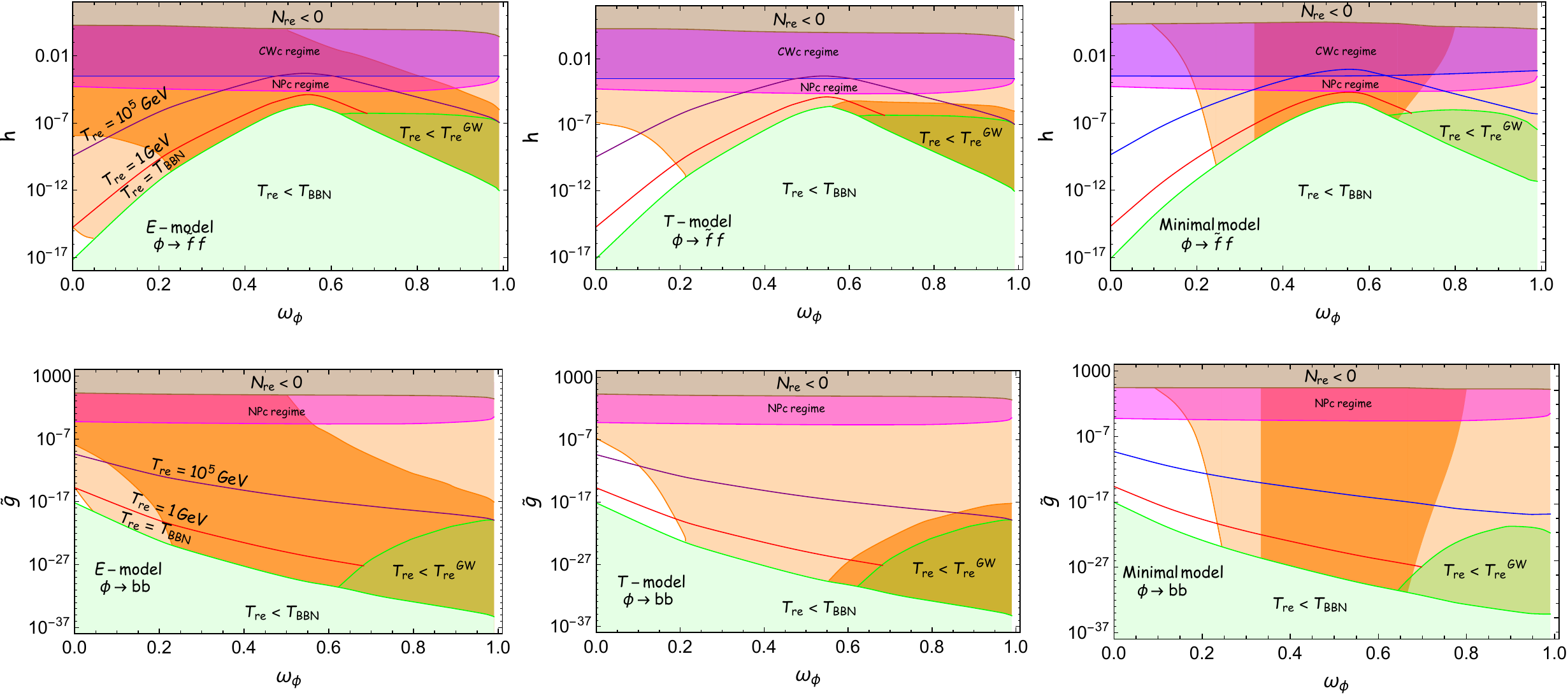}
          \caption{\em The impact of {\rm PLANCK 2018 + BICEP/Keck + PGW + BBN + CWc + NPc} on dimensionless inflaton-radiation coupling (h,$\tilde{g}$) with respect to $w_{\phi}$ for three different models. The color codes are the same as before. For these plots, we have taken ($\alpha,\,\phi_*)=10$.}
          \label{couplingw}
          \end{center}
      \end{figure*}
As an example for $w_\phi=0$, all reheating temperature above $T_{\rm re}>10^{10}$ GeV lies in the nonperturbative regime. This nonperturbative bound is sensitive to both $(\alpha,w_\phi)$; for better visualization, see, Figs.\ref{Tree}, \ref{Tret} and details numerical  values given in Tables-\ref{rehconE}, \ref{rehconT}.\\
For a complete understanding of the reheating parameter space, finally, in Fig.\ref{couplingw}, we plotted the parametric dependence of inflaton-radiation coupling with respect to reheating equation of state. The left two column plots are for E and T-models corresponding to $\alpha = 10$. Interestingly, if one confines within the $1\sigma$ region of PLANCK, for T-model inflaton-radiation couplings are tightly bounded. As an example, for $w_{\phi} =0.82$, one obtains $10^{-7} > h > 10^{-6}$, and 
$10^{-23} > \tilde{g} > 10^{-22}$.
\subsection {Minimal model and constraints}
Detailed phenomenological constraints of the minimal inflation can be observed from Figs.\ref{nsrminimal}, \ref{w13} and \ref{Trem}, and are summarized in Tables-\ref{infresE0}, \ref{minimalinf}, \ref{minimalres}. CWc sets the upper limit on inflaton-radiation coupling as $\tilde{g}_{\rm max}\sim \mathcal{O}(10^{-1})$ and $h_{\rm max}\sim \mathcal{O}(10^{-4})$. These available bounds on couplings thereafter set restrict $T_{\rm re}^{\rm max}\sim 10^{15}$ GeV for bosonic reheating, which is close to instantaneous reheating. However, for inflaton-fermion coupling, $T_{\rm re}^{\rm max}$ is noticeably sensitive to the inflaton parameter $(\phi_*,\,w_\phi)$ value and detailed numerical values are provided Table-\ref{minimalres} and Fig.\ref{Trem}.\\
Unlike attractor models, the lower bound on $T_{\rm re}$ for this model has non-trivial dependence on the potential parameter $\phi_*$. For a given equation of state, one observes that there exists a critical value of $\phi_*$ above which value lower bound decreases from $T_{\rm re}^{\rm GW}$ to $T_{\rm BBN}$. For instance in the first plot of Fig.\ref{Trem}, one clearly see that for $\phi_*>155$, $T_{\rm re}^{\rm min}\sim 10^4$ GeV drops to $T_{\rm BBN}$. This essentially suggests that as on increase $\phi_*$, the lower possible inflation-fermion coupling decreases from $h_{\rm min} \sim 10^{-6}$ to $10^{-10}$. In Table-\ref{minimalres}, we provide maximum and minimum values of coupling parameters and reheating temperature for different EoS $w_\phi=(0,\,1/3,\,0.67,\,0.82,\,0.99)$. For specific model parameters, bounds on the reheating parameters will always be within the bounds provided in Table-\ref{minimalres}.\\
If one considers the nonperturbative bound, the upper limit on the coupling parameter is further modified (see Fig.\ref{Trem}). Additionally, it depicts the fact that compared to the bosonic coupling $\tilde{g}$ nonperturbative bounds on the fermionic coupling $h$ is much more sensitive to $\phi_*$ when $w_\phi>1/3$. Finally, in Fig.\ref{couplingw}, the right column figures depict the parametric variation of inflaton-radiation coupling with respect to the $w_{\phi}$ for the minimal model with $\phi_* = 10$. Interestingly, the minimal model is non-viable for the lower equation of state dictated by PLANCK together with BICEP/$Keck$ data, which corresponds to white regions. This can also be observed from Fig.\ref{Trem}. As an example, $2\sigma$ ($n_s-r$) bound ruled out $\phi_* \gtrsim 9.4$, whereas if we consider $1\sigma$ bound ruled out $\phi_* \gtrsim 5.5$.


\section{Conclusions}\label{sc8}
In this paper, we have analyzed in detail the phenomenological implications of the latest PLANCK 2018 and BICEP/Keck data on the physics of inflaton. We analyzed three different plateau inflation models along with two different reheating scenarios. We particularly discussed two different channels of inflaton decaying into fermions through $\phi \to \bar f f$ and bosons through $\phi \to  bb$. Along with the observations, we further employ bound from PGWs and both perturbative and nonperturbative constraints on the inflaton-radiation coupling.

At the first step, considering the allowed range of reheating temperature ($T_{\rm BBN} \,(\,T_{\rm re}^{\rm GW}\,), 10^{15}\,\rm GeV$), we project our result in the ($n_s-r$) plane to fix the upper limit on the potential parameter $\alpha(\phi_*)$ at $1\sigma$ and $2\sigma$ C.L. Once we fixed $\alpha(\phi_*)$ for five different sets of inflaton equation of state $w_\phi=(0,\,1/3,\,0.67,\,0.82,\,0.99)$, that in turn fixes inflationary parameters such as the range of $N_k$ and $H_{\rm end}$. The limiting values of inflationary parameters for different models  are shown in Tables-\ref{infresE}, \ref{infresE0}, \ref{minimalinf}. \\
In the second step, considering those bounds on inflationary parameters, we analyze the restriction on the reheating parameters such as $T_{\rm re}$ and couplings $\,h,\tilde{g}$. The upper bound on inflaton-radiation coupling $h, \tilde{g} \sim 0.1$, is derived from the $T_{\rm re}^{\rm max}\sim 10^{15}$ GeV value corresponding to instantaneous reheating though a simple relation $T_{\rm re} \propto \sqrt{\Gamma_{\phi}}$. However, once CWc and NPc are employed, the upper bound of coupling significantly drifted from the aforementioned value. The detailed quantitative values of different bounds arising from the theory constraints are displayed in Tables \ref{rehconE}, \ref{rehconT}, and \ref{minimalres} for three different plateau type inflationary model attractor E, T, and minimal model respectively. The lower bound on reheating temperature is conventionally fixed by $T_{\rm BBN}$ (deep green shaded region in all the plots). However, if we incorporate PLANCK + BICEP/Keck + PGWs constraints, such a possibility is no longer generic. If $w_{\phi} < 1/3$, in most of the parameter region for the E and T model, $T_{\rm BBN}$ lies outside the $1\sigma$ region (see fourth column of Fig.\ref{Tree} and \ref{Tret}). However, for those models PGWs bound on $T_{\rm re}^{\rm min} = T_{\rm re}^{\rm GW}$ is large by  several decade above $T_{\rm BBN}$ for $w_{\phi} > 0.6$. As an example for $w_{\phi} =0.99$, $T_{\rm re}^{\rm min} = T_{\rm re}^{\rm GW} \sim 10^{5}$ GeV. 
For the minimal model, however, the lower bound on reheating temperature is noticeably  dependent on $\phi_*$ specifically for the higher equation of state. 

Apart from direct and indirect observational bounds, we further utilized the theoretical bounds coming from CWc and NPc. CWc strictly restricts the maximum allowed value of $(h,g)$, above which inflation will be jeopardized. However, if we assume the fact that reheating process itself is perturbative in nature, it is NPc that plays a wider role in setting the upper limit on both the reheating temperature and inflaton-radiation couplings for all the models under consideration. If nature chooses the coupling to be within the nonperturbative regime, some of the present conclusions may be affected, which we defer for our future study.

To the end, we wish to note that future experiments such as CMB-S4 \cite{CMB-S4:2020lpa} and LiteBIRD \cite{LiteBIRD:2022cnt} will be able to put stronger bound in $(n_s, r)$ plane which will certainly improve our understanding the precise nature of beyond standard model physics such as inflaton (see, for instance, the recent works \cite{Drewes:2022nhu,Drewes:2023bbs}). 
\section{acknowledgments}
R.M. would like to thank the Ministry of Human Resource Development, Government of India (GoI), for financial assistance. M.R.H wishes to acknowledge support from the Science and Engineering Research Board~(SERB), Government of India~(GoI), for the National Post-Doctoral fellowship, File Number: PDF/2022/002988.DM wishes to acknowledge support from the Science and Engineering Research Board~(SERB), Department of Science and Technology~(DST), Government of India~(GoI), through the Core Research Grant CRG/2020/003664.
\appendix
\hspace{0.5cm}
 
\end{document}